%% file: main.tex
\documentclass[pdflatex,sn-mathphys-num]{sn-jnl}% Math and Physical Sciences Numbered Reference Style 
\usepackage{graphicx}%
\usepackage{multirow}%
\usepackage{amsmath,amssymb,amsfonts}%
\usepackage{amsthm}%
\usepackage{mathrsfs}%
\usepackage[title]{appendix}%
\usepackage{xcolor}%
\usepackage{textcomp}%
\usepackage{manyfoot}%
\usepackage{booktabs}%
\usepackage{algorithm}%
\usepackage{algorithmicx}%
\usepackage{algpseudocode}%
\usepackage{listings}%
\usepackage{makecell}
\theoremstyle{thmstyleone}%
\theoremstyle{thmstyletwo}%

\theoremstyle{thmstylethree}%

\begin{document}

\title[Article Title]{CrysToGraph: A Comprehensive Predictive Model for Crystal Materials Properties and the Benchmark}

%%=============================================================%%
%% GivenName	-> \fnm{Joergen W.}
%% Particle	-> \spfx{van der} -> surname prefix
%% FamilyName	-> \sur{Ploeg}
%% Suffix	-> \sfx{IV}
%% \author*[1,2]{\fnm{Joergen W.} \spfx{van der} \sur{Ploeg} 
%%  \sfx{IV}}\email{iauthor@gmail.com}
%%=============================================================%%

\author*[1,5]{\fnm{Hongyi} \sur{Wang}}\email{howard.wang@my.cityu.edu.hk}

\author[2]{\fnm{Ji} \sur{Sun}}
\author[1]{\fnm{Jinzhe} \sur{Liang}}
\author[1]{\fnm{Li} \sur{Zhai}}
\author[3]{\fnm{Zitian} \sur{Tang}}
\author[1]{\fnm{Zijian} \sur{Li}}
\author[1]{\fnm{Wei} \sur{Zhai}}

\author[4]{\fnm{Xusheng} \sur{Wang}}
\author[5]{\fnm{Weihao} \sur{Gao}}
\author*[6]{\fnm{Sheng} \sur{Gong}}\email{sheng.gong@bytedance.com}

\affil*[1]{\orgdiv{Department of Chemistry}, \orgname{City University of Hong Kong}, \orgaddress{\city{Kowloon}, \state{Hong Kong},  \country{China}}}
\affil[2]{\orgdiv{School of Mathematics}, \orgname{Renmin University of China}, \orgaddress{\state{Beijing}, \country{China}}}
\affil[3]{\orgdiv{Thrust of Data Science and Analytics}, \orgname{, The Hong Kong University of science and technology (Guangzhou)}, \orgaddress{\city{Guangzhou}, \state{Guangdong}, \country{China}}}
\affil[4]{\orgdiv{School of Pharmacy}, \orgname{Guangxi University of Chinese Medicine}, \orgaddress{\city{Nanning}, \state{Guangxi}, \country{China}}}
\affil[5]{\orgname{ByteDance Research}, \orgaddress{\state{Beijing}, \country{China}}}
\affil[6]{\orgname{ByteDance Research}, \orgaddress{\state{Washington}, \country{USA}}}
%%==================================%%
%% Sample for unstructured abstract %%
%%==================================%%

\abstract{The bonding across the lattice and ordered structures endow crystals with unique symmetry and determine their macroscopic properties. Crystals with unique properties such as low-dimensional materials, metal-organic frameworks and defected crystals, in particular, exhibit different structures from bulk crystals and possess exotic physical properties, making them intriguing subjects for investigation. To accurately predict the physical and chemical properties of crystals, it is crucial to consider long-range orders. While GNN excels at capturing the local environment of atoms in crystals, they often face challenges in effectively capturing longer-ranged interactions due to their limited depth. In this paper, we propose CrysToGraph ($\textbf{Crys}$tals with $\textbf{T}$ransformers $\textbf{o}$n $\textbf{Graph}$s), a novel and robust transformer-based geometric graph network designed for unconventional crystalline systems, and UnconvBench, a comprehensive benchmark to evaluate models' predictive performance on multiple categories of crystal materials. CrysToGraph effectively captures short-range interactions with transformer-based graph convolution blocks as well as long-range interactions with graph-wise transformer blocks. CrysToGraph proofs its effectiveness in modelling all types of crystal materials in multiple tasks, and moreover, it outperforms most existing methods, achieving new state-of-the-art results on two benchmarks. This work enhances the development of novel crystal materials in various fields, including the anodes, cathodes and solid-state electrolytes.}


%%\pacs[JEL Classification]{D8, H51}

%%\pacs[MSC Classification]{35A01, 65L10, 65L12, 65L20, 65L70}

\maketitle

\input{main/intro}

\input{main/methods}

\input{main/results}

\input{main/discussion}

\backmatter

\input{main/back}

%%===========================================================================================%%
%% If you are submitting to one of the Nature Portfolio journals, using the eJP submission   %%
%% system, please include the references within the manuscript file itself. You may do this  %%
%% by copying the reference list from your .bbl file, paste it into the main manuscript .tex %%
%% file, and delete the associated \verb+\bibliography+ commands.                            %%
%%===========================================================================================%%

\bibliography{sn-article}% common bib file
%% if required, the content of .bbl file can be included here once bbl is generated
%%\input sn-article.bbl

\end{document}

%% file: main/intro.tex
\section{Introduction}\label{sec1}

Graph Neural Networks (GNNs) represent a significant breakthrough in the field of machine learning when applied to graph-structured data. These networks are extremely appropriate for handling data that can be organized as topological graphs. Such graph structures are prevalent in real-world scenarios, including knowledge graphs \citep{nathani2019knowledge2, hamaguchi2017knowledge3, schlichtkrull2018knowledge4, wang2018knowledge5}, social networks \citep{zhang2018social1, qiu2018social2, liu2019social3}, recommendation systems \citep{ying2018recomm1, monti2017recomm2, berg2017recomm3}, and also, natural science\citep{santoro2017phy1, battaglia2016phy2, duvenaud2015chem1, fout2017bio1}. GNNs have also found great success in modeling small molecules \citep{kearnes2016chem2, pei2023graph, cai2022midruglikeness, zhu2022neural, gong2024bamboo}. Molecules, with atoms connected by covalent bonds, can be depicted as graphs naturally. This success with small molecules extends to related fields, such as inorganic crystals \citep{gong2022examining} and biological macromolecules \citep{zhang2022protein}. Covalent bonds and Coulomb interactions are the primary forces responsible for packing atoms into crystals. GNNs are successful in capturing these short-range interactions. Yet, constrained by their depth, GNNs focus mainly on the local environment and struggle to capture global information in graphs. Hence, capturing information such as long-range orders which is crucial in crystalline systems is challenging for GNNs.

The history of application of GNNs on crystal structures dates back to CGCNN \citep{xie2018cgcnn} was the first one developed primarily focused on crystalline structures. CGCNN incorporated geometric construction of periodic multi-graphs and adopted a message-passing approach that concatenateed node features from central and neighboring nodes, along with corresponding edge features. Subsequent models include iCGCNN \citep{park2020icgcnn} which introduced Voronoi structures \citep{lee1982voronoi} for modeling three-body relation, GeoCGNN \citep{cheng2021geocgnn} which utilized attention masks and plane waves to encode local geometrical information, and MEGNet \citep{chen2019megnet} which incorporated global state information and edge updates. Further, ALIGNN \citep{choudhary2021alignn} introduced line graphs to model geometric connectivity, while coGN and coNGN \citep{ruff2023cogn} introduced nested line graphs to explicitly model higher-ordered connectivity information. Also, Matformer \citep{yan2022matformer} and Comformer \citep{yan2024complete} utilized attention mechanism with periodic pattern encodings in modeling crystalline systems. Other techniques like contrastive learning \citep{bai2023xtal2dos, kong2022xtal2dos2} and prototypical classifiers \citep{bai2019imitation} are also applied in the training of GNNs for crystals. Numerous studies have been conducted on forecasting crystal characteristics, with a predominant emphasis on the local surroundings and conventional materials. Previous works have achieved great success on the benchmarks of traditional crystals, while limited attention has been directed towards many specific types of material and the long-range order present within crystals \citep{chen2019megnet, de2021modnet}. Furthermore, the existing benchmarks for predicting properties of crystal materials  \citep{dunn2020matbench, jain2013materialsproject, choudhary2020jarvis} drew limited attention to certain types of materials, for example, the 2D materials which often demonstrate unique electronic properties due to the ultra-thin layered structures, the MOFs with unique adsorption and catalytic properties endowed by the porous structure, and defected crystals with potential catalytic capabilities.

\begin{figure*}[!]
\begin{center}
\includegraphics[height=6cm]{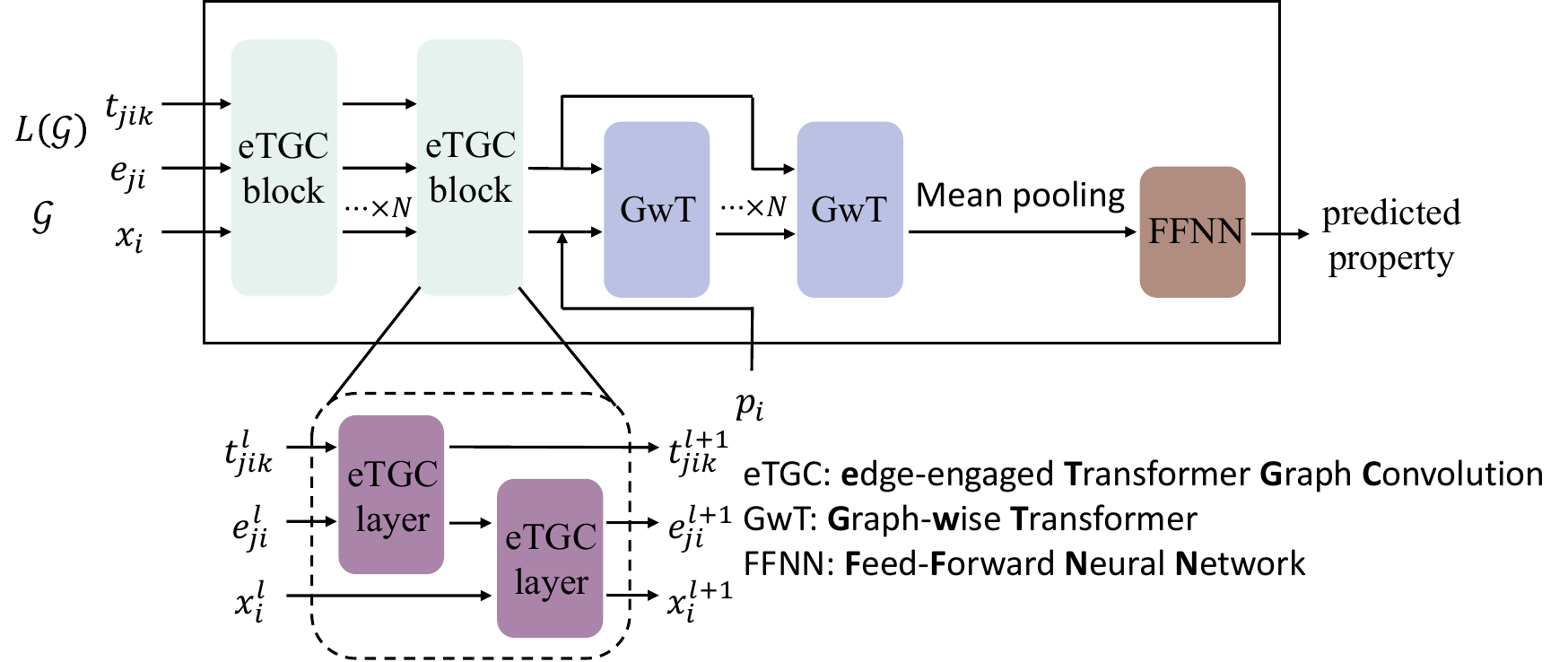}
\end{center}
\caption{An overview of the architecture of CrysToGraph. In this paper, $\mathcal{G}$ denotes the original crystal graph, $L(\mathcal{G})$ denotes the line graph built upon the edges of the direct crystal graphs. For inputs, $x_i$ denotes the atom (node) feature of node $i$, $e_{ji}$ represents the bond (edge) feature of edge $(j,i)$, $t_{jik}$ represents the edge feature in the line graph, also the relationship between edge $(j,i)$ and $(i,k)$, $p_i$ denotes the positional encoding on atom $i$. Details of the graphs and positional encoding can be found in section 3.1 and 3.2. Details of the architecture can be found in section 3.3.}\label{fig_overview}
\end{figure*}

In this work, we present CrysToGraph, a transformer-based geometric graph network designed for crystalline systems. CrysToGraph employs a novel architecture that combines transformer-based message passing blocks for updating node and edge features with a graph-wise transformer for explicitly incorporating both local environments and long-range interactions. Using line graphs to explicitly engage the geometric and connectivity information, CrysToGraph effectively captures the information in complex crystal materials including 2D materials, Metal-organic frameworks (MOF) and defected crystals, displaying a high potential in discovering novel materials. CrysToGraph outperforms most models on datasets of crystal materials, achieving state-of-the-art results in 10 datasets out of 15 and establishing it as one of the best models for predicting crystal material properties. We also present UnconvBench, a benchmark with 11 datasets to comprehensively evaluate the models' performance on specific types of crystal materials. We summarize our main contributions as follows:
\begin{enumerate}
    \item We propose \textit{CrysToGraph}, a transformer-based geometric graph network for explicitly capturing short-range and long-range interactions in crystalline systems, as shown in Figure~\ref{fig_overview}.
    \item We propose \textit{eTGC} (edge-engaged transformer graph convolution), a transformer-based graph convolution layer that updates node features and edge features using a shared attention score calculated based on the features of the central node, neighboring nodes and edges.
    \item We propose \textit{GwT} (graph-wise transformer), a transformer encoder tailored for graphs, to capture long-range dependencies among nodes on a graph-wide scale.
    \item We present UnconvBench, a benchmark with 10 datasets in bulk crystals, 2D crystals, MOFs and defected crystals to comprehensively evaluate the performance of machine learning models in modelling crystals of different size, dimension, and symmetry. 
\end{enumerate}

%% file: main/methods.tex
\section{Methods}\label{sec3}

\subsection{Crystal Graphs}

The graphs are constructed from the structure of crystals. For every crystal, we construct a crystal graph with atoms as nodes. Edges between the nodes are identified in a k-nearest-neighbors manner. Line graphs are constructed based on the crystal graphs to explicitly model connectivity and three-body interactions within the crystals.

\subsubsection{Nodes}

Node features play a pivotal role in GNNs, and they represent essential information pertaining to the atoms in a crystal when applied to crystal structures. Each atom within the crystal is depicted as a node in the graph, with corresponding embeddings. Our atom embeddings inherited a set of CGCNN-style atom embeddings of 92 dimensions. The CGCNN atom embeddings are a curated set of features generated in one-hot encoding scheme. This encoding encapsulates various atomic properties of the atom represented by the node. These embeddings are atom-specific and do not inherently account for neighboring information or atomic charges.

\begin{figure}[h]
\begin{center}
%\framebox[4.0in]{$\;$}
\includegraphics[height=3cm]{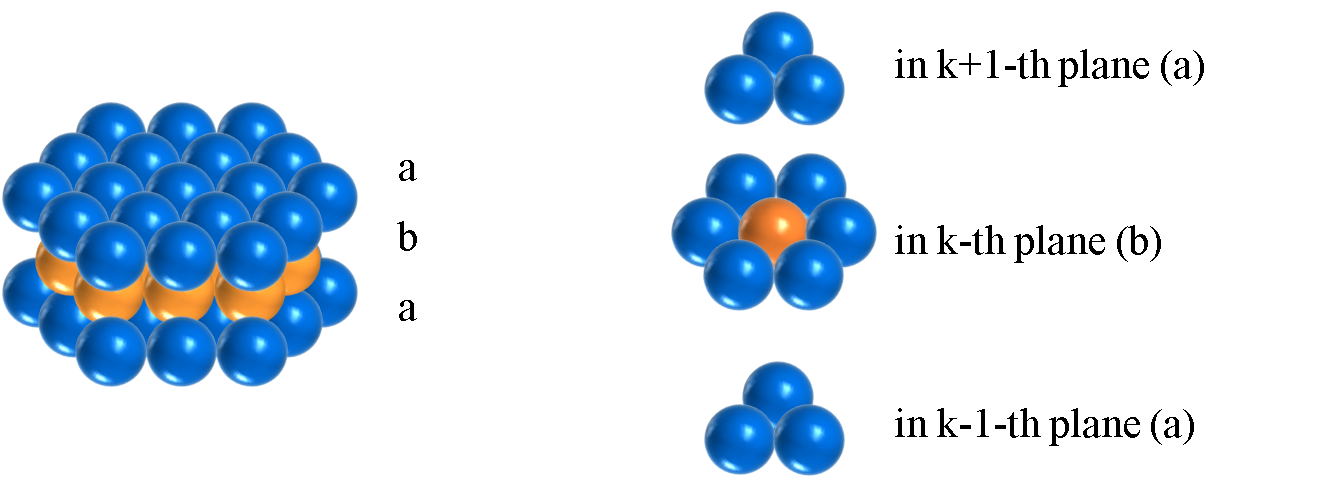}
\end{center}
\caption{Left: A hexagonal close-packed ($hcp$) structure with atoms stacked in layers in an \textit{abab\dots} pattern into a bulk crystal. Right: The theoretical maximum number of neighbors for a single atom (illustrated in orange) is 12.}\label{fig_hcp}
\end{figure}

\subsubsection{Edges}

Edges in crystal graphs represent the bonds connecting atoms and are a fundamental aspect of inorganic crystal structures. The number of edges per node, or atom coordination number in the context of crystallography, typically varies between 2 and 12 in crystal structures \citep{wells2012inorganic}. In the densest type of crystal structure, such as hexagonal close-packed ($hcp$) and cubic close-packed ($ccp$), each atom's coordination number reaches the theoretical maximum of 12, as shown in Figure~\ref{fig_hcp}.

In our study, we apply the k nearest neighbor (k-NN) method to identify the edges around nodes, setting the value of k to 12. This choice aligns with the theoretical maximum number of neighbors and aims to maximize the incorporation of neighbor information in each message-passing step. Importantly, it should be noted that, here, 12 nearest neighbors are considered connected, even if their physical distance may extend up to 20 \AA.

To calculate edge features, we use the shifts in nodes' positions represented using spherical coordinates. These position shifts are expanded using radial-based filters to increase the dimension non-linearly. Additionally, we introduce a Boolean term in the edge features to indicate whether the distance between two neighbors exceeds the ion bond length cutoff. In this work, we define the threshold for the longest ion bond as 8 \AA.

\subsubsection{Line graphs}

\begin{figure}[t]
\begin{center}
%\framebox[4.0in]{$\;$}
\includegraphics[height=2cm]{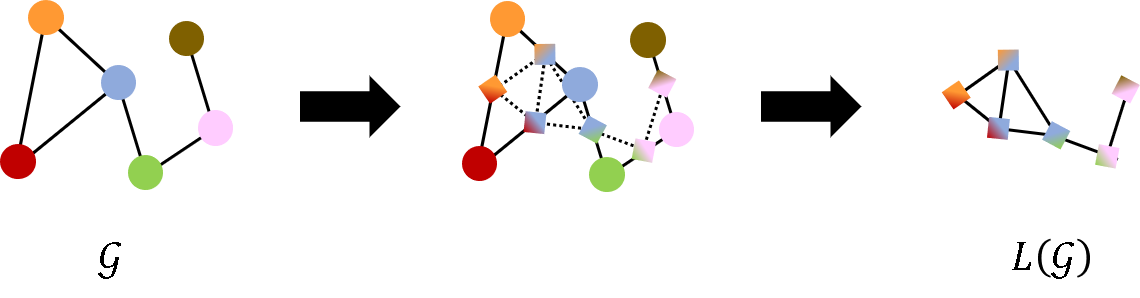}
\end{center}
\caption{Construction of line graph $L(\mathcal{G})$ from direct graph $\mathcal{G}$. The edges in direct graph $\mathcal{G}$ are considered as nodes in line graph $L\mathcal{G}$, and the angles of two edges in $\mathcal{G}$ are constructed as the edges in $L\mathcal{G}$.}\label{fig_line}
\end{figure}

The line graph $\displaystyle L(\mathcal{G})$ of a given graph $\displaystyle \mathcal{G} $ is a graph where the nodes in $\displaystyle L(\mathcal{G})$ represent the edges in $\mathcal{G}$, and the edges in $\displaystyle L( \mathcal{G}) $ correspond to edges pairs in $\displaystyle \mathcal{G} $, as illustrated in Figure~\ref{fig_line}. Specifically, for any pair of edges $\displaystyle (n_u, n_v) $ and $\displaystyle (n_v, n_w) $ in $\displaystyle \mathcal{G} $, there exist corresponding nodes $\displaystyle e_u$ and $\displaystyle e_v$ in the line graph $\displaystyle L(\mathcal{G}) $. Moreover, there is an edge $\displaystyle (e_u, e_v)$ in $\displaystyle L(\mathcal{G})$, and the features of this edge are derived by expanding using radial-based filters based on the cosine of the angle between the edges $\displaystyle (n_u, n_v)$ and $\displaystyle (n_v, n_w)$ in $\mathcal{G}$.

In essence, the line graph provides a higher-level representation where edges in the original graph become nodes, and connections in the line graph signify relationships between pairs of edges in the original graph, capturing information about their angles and structural configurations.

\subsection{Positional Encoding of Atoms}

Conventional GNNs do not require positional encodings because they operate on graph structures that lack spatial information but inherently embody connectivity. However, in our work, we devise a graph-wise transformer structure to capture the long-range interactions, making positional encoding indispensable. Inspired by previous studies \citep{gao2022learning, gao2022supervised}, this structure incorporates positional encodings to effectively process spatial information.

To address this requirement, we employed a comprehensive approach to positional encodings that combined multiple sources of positional information, including:

\begin{enumerate}
\item Laplacian positional encoding \citep{dwivedi2020lappe}: a positional encoding based on the Laplacian operator, which captures structural relationships within the graph.
\item Random walk positional encoding \citep{dwivedi2021rwpe}: a positional encoding derived from random walk processes, providing additional information about connectivity.
\end{enumerate}

By concatenating these positional encodings, we aim to encode the representation of spatial and connectivity information within the graph. Moreover, this comprehensive approach ensured that positional encodings captured both the absolute and relational aspects of node positions and the connectivity, facilitating the effectiveness of the subsequent graph-wise transformers in our model.

\subsection{Model Architecture}

Here, we introduce the architecture of CrysToGraph. The model contains 3 parts: edge-engaged transformer graph convolution for modeling short-range interactions, graph-wise transformers for modeling long-range interactions and feed forward linear layers for predicting of task-specific properties. The input crystal graphs consist of the direct graphs and line graphs, and the outputs of the entire model are the properties of the crystal. The detailed structure is shown in Figure~\ref{fig_overview}. To improve the accurccy of prediction, ensembles of model with mean of all outputs as the final output can be applied.

The CrysToGraph model can be defined as:

\begin{equation}
% \begin{center}
    CrysToGraph(\mathcal{G}, L(\mathcal{G})) = FFNN(GwT_{\times N}(eTGC_{\times N}(\mathcal{G}, L(\mathcal{G}))) \label{eqo1}
% \end{center}    
\end{equation}

\subsubsection{Edge-engaged Transformer Graph Convolution (eTGC)}

\begin{figure}[h]
    \begin{center}
        \includegraphics[height=8.5cm]{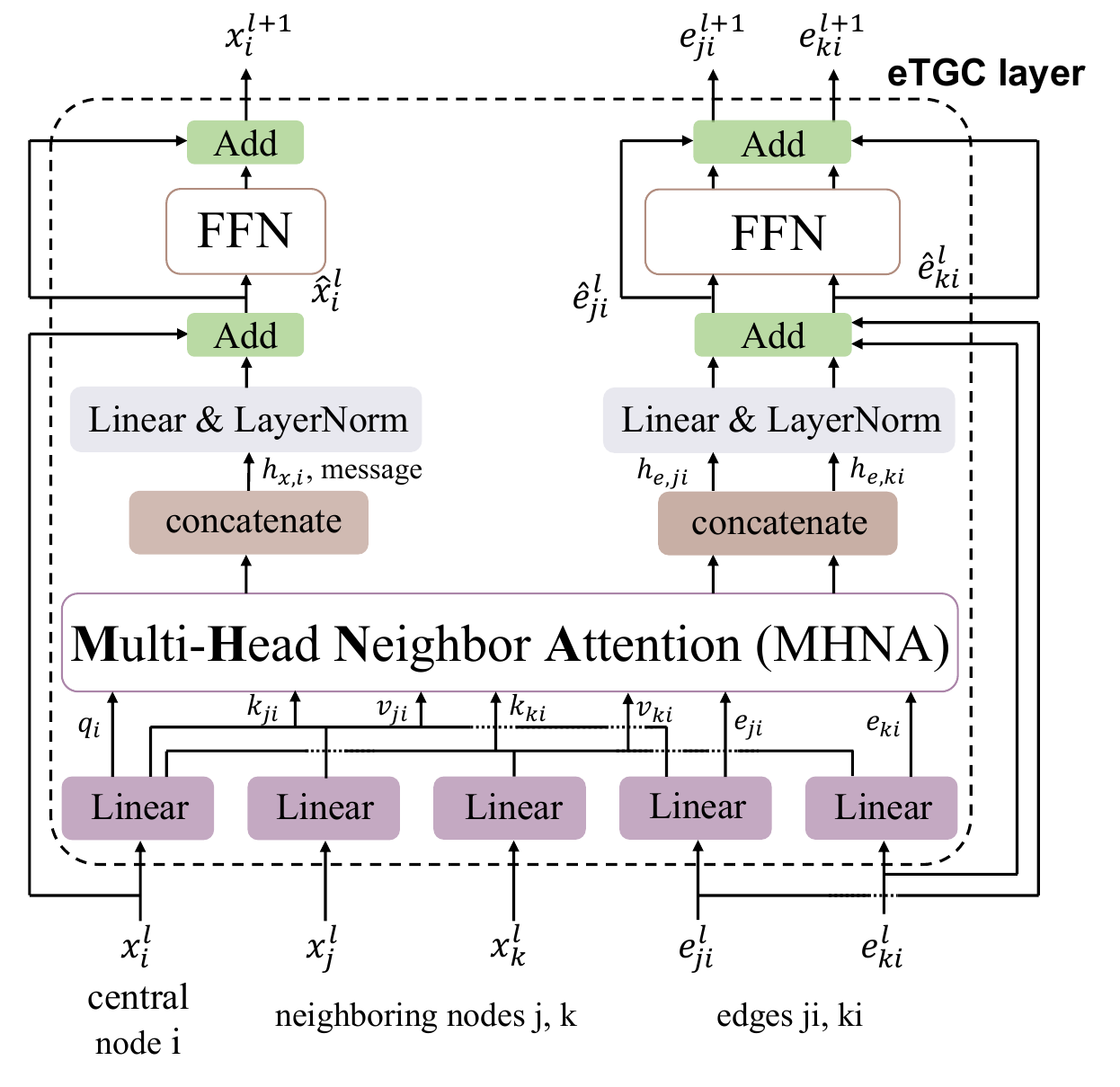}
    \end{center}
\caption{Structure of an eTGC layer. An eTGC block consists of two eTGC layers that take different inputs.}\label{fig_etgc}
\end{figure}

\begin{figure}
\begin{center}
\includegraphics[height=4.2cm]{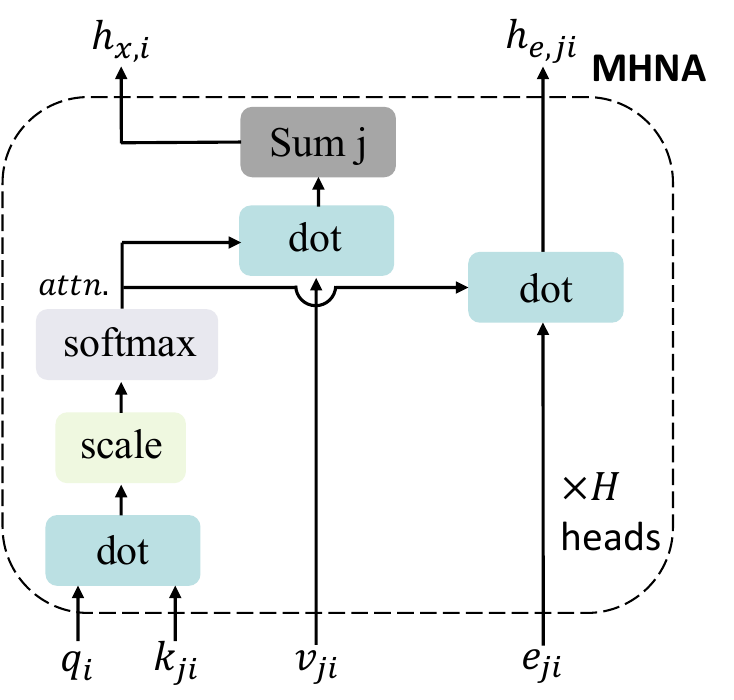}
\end{center}
\caption{Structure of multi-head neighbor attention in eTGC layers.}\label{fig_mhna}
\end{figure}

\begin{figure}
\begin{center}
\includegraphics[height=4.2cm]{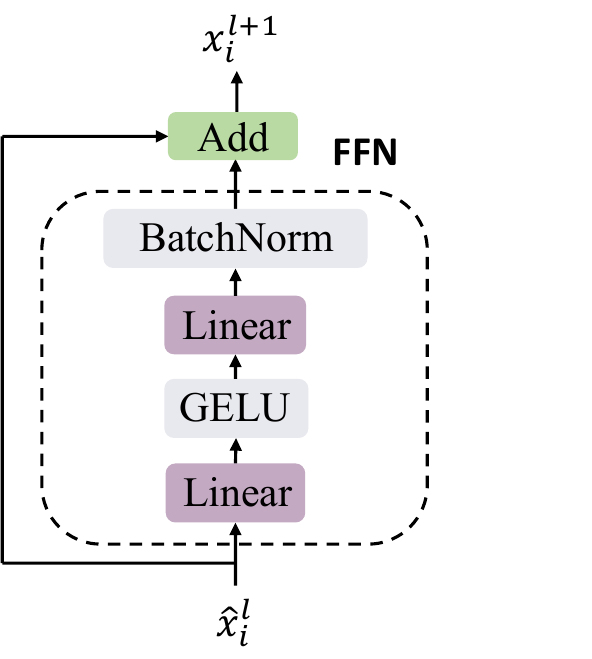}
\end{center}
\caption{Structure of FFN in eTGC layers and GwT layers.}\label{fig_ffn}
\end{figure}

Each eTGC block consists of two eTGC layers for direct graph and line graph, respectively. Node features, edge features and edge features of the line graphs are updated in each block. The eTGC layer for the line graph updates the edge features of the direct graph and the edge features of the line graph first, then the eTGC layer for direct graph updates the node features and edge features of the direct graph. As demonstrated in Figure~\ref{fig_etgc}, Figure~\ref{fig_mhna}, and Figure~\ref{fig_ffn} an eTGC layer contains linear transformation, multi-head neighbor-attention (MHNA) and a feed-forward network (FFN). 

In each eTGC layer, input node features are linearly transformed into query, key and value vectors $Q$, $K$ and $V$. The edge features are linear transformed into vector $E’$:

\begin{equation}
    Q = W_qX, 
    K = W_kX,
    V = W_vX,
    E' = W_eE \label{eqe1}
\end{equation}

Scoping to a single central node and its neighbors, $k$ and $v$ of both the central node and the neighboring nodes are concatenated with edge feature $e'$ in a CGCNN manner. The neighbor-attention is calculated using $q_i$ of central node and linear transformation results of the concatenated vectors $k_{ji}$ and $v_{ji}$. Node features and edge features are updated with shared attention scores which are scaled with constant $\sqrt{d_k}$ and softmax layer. For simplicity of illustration, we consider a single-head attention and assume $d_k = d_v$ in the equations:

\begin{equation}
k_{ji} = W_{ke}(k_i,k_j,e'_{ji}),
v_{ji} = W_{ve}(v_i,v_j,e'_{ji}) \label{eqe2}\\
\end{equation}
\begin{equation}
attn = softmax(\frac{q_ik_{ji}}{\sqrt{d_k}}) \label{eqe3}\\
\end{equation}
\begin{equation}
h_{x,i}=\sum attn\cdot v_{ji},
h_{e,ji} = attn\cdot e'_{ji} \label{eqe4}\\
\end{equation}

where $h_x$ and $h_e$ denote the hidden outputs of the multi-head neighbor-attention. Outputs of all heads are concatenated and linear transformed into hidden outputs, among which the hidden output $h_{x,i}$ is the message being passed to a certain node $i$ .

Scoping back to the graph, hidden outputs of the multi-head neighbor-attention are linearly transformed and layer normalized, before being added to the input central node features or edge features:

\begin{equation}
{\hat{x}}^l = x^{l} + LayerNorm(W_{on}{h_x}^l) \label{eqe5}\\
\end{equation}
\begin{equation}
{\hat{e_{ji}}}^l = {e_{ji}}^{l} + LayerNorm(W_{oe}{h_e}^l) \label{eqe6}
\end{equation}

where $\hat{x}$ and $\hat{e}$ denote the node outputs and edge outputs of the multi-head neighbor-attention at each node and edge, respectively.

Each multi-head neighbor-attention sublayer is followed by a feed-foward network sublayer in an eTGC layer. The FFN contains a batch normalization and two linear functions with GELU as activation function in the middle. A residual connection is applied at the end:

\begin{equation}
x^{l+1} = {\hat{x}}^{l} + BatchNorm(W_{x2}Gelu(W_{x1}{\hat{x}}^l + b_{x1})+b_{x2}) \label{eqe7}\\
\end{equation}
\begin{equation}
e^{l+1} = {\hat{e}}^{l} + BatchNorm(W_{e2}Gelu(W_{e1}{\hat{e}}^l + b_{e1})+b_{e2}) \label{eqe8}\\
\end{equation}

where  
\begin{equation}
    Gelu(x) = xP(X<x) = 0.5x(1+erf(\frac{x}{\sqrt(2})) \label{eqe9}
\end{equation}.

\subsubsection{Graph-wise Transformer (GwT)}

\begin{figure}[h]
    \begin{center}
        \includegraphics[height=11cm]{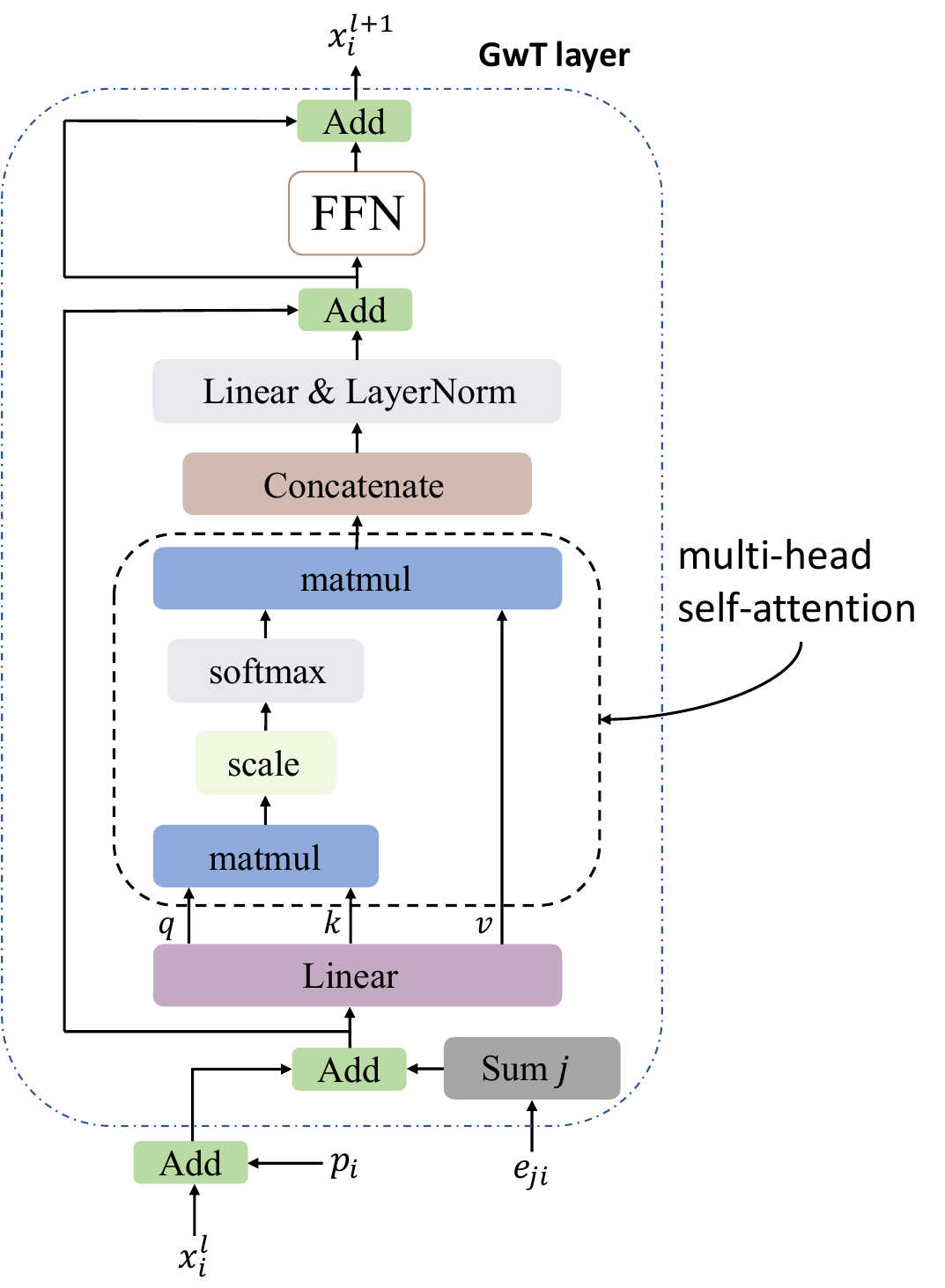}
    \end{center}
\caption{Structure of a GwT layer. A graph-wise multi-head self attention is the core block of this layer which grants the layer a fully-connected nature to model the interaction within the graph.}\label{fig_gwt}
\end{figure}

Chemically, long-range order distinguishes crystals from small molecules by providing them with a regular structure, macroscopic symmetry, and unique optical properties, while short-range interactions are responsible for stabilizing the crystal structure on a local scale. In addition to the eTGC blocks, which focus on short-range interactions at the local scale, we implemented a GwT layer, of which the structure is shown in Figure~\ref{fig_gwt}, to explicitly model the long-range interactions within the entire crystal. 

In this GwT layer, no connectivity information is explicitly incorporated since all nodes in each graph are treated as an ordered sequence of tokens. Thus, positional encodings and the sum of incoming edge features are added to the node features in prior:

\begin{equation}
x_{i,pe} = x_i + W_{pe}p_i + \sum_j W_ee_{ji} \label{eqg1}
\end{equation}
where $p_i$ denotes the positional encoding of node $i$.

Typical multi-head self-attentions mechanism is applied on node features in the entire graphs, followed by a residual feed forward network with layer normalization:

\begin{equation}
  Q=W_qX, K=W_kX, V=W_vX \label{eqg2}\\
\end{equation}
\begin{equation}
  Attn=softmax(\frac{QK^T}{\sqrt{d_k}}) \label{eqg3}\\
\end{equation}
\begin{equation}
  H^l=X^{l}+LayerNorm(W_o(Attn\cdot V^{l})) \label{eqg4}\\
\end{equation}
\begin{equation}
  X^{l+1} = H^l + LayerNorm(W_2Gelu(W_1H^l+b_1)+b_2) \label{eqg5}
\end{equation}

\subsubsection{Feed-Forward Neural Network (FFNN)}

To capture both short-range and long-range interactions within the crystals, we employed eTGC blocks and GwT. These components model interactions at different scales. Each graph's node features are aggregated by taking their mean, resulting in a graph-level feature. This feature is then layer-normalized and fed into a task-specific FFNN for predicting specific properties. The task-specific feed-forward neural network we implemented is a multi-layer perceptron, utilizing Softplus as its activation function:

\begin{equation}
    X^l = Softplus(W_lX^{l-1}+b_l) \label{eqf1} \\
\end{equation}
\begin{equation}
    X^{out} = W_nX^{n-1} + b_n \label{eqf2}
\end{equation}

where $X^{out}$ denotes the final prediction if the task is a regression task. For classification tasks, positive $X^{out}$ yield positive predictions, while negative $X^{out}$ yield negative predictions.

%% file: main/results.tex
\section{Results}\label{sec2}

Experiments were conducted on an assembly of predictive tasks that explain the functionality of each parts of the CrysToGraph and the effectiveness of the model as a whole. The raw data are collected from Materials Project \citep{jain2013materialsproject} and JARVIS \citep{choudhary2020jarvis}. In this section, we present the UnconvBench and evaluate our model from 5 aspects: exploration of model structure, performance on large-cell crystals, performance on defected crystals, performance on UnconvBench and performance on traditional crystal benchmark.

\subsection{Benchmark}

The benchmark for evaluating models' performance on s crystal materials comprises a curated set of tasks with distinct targets. The samples in this benchmark encompass a broad spectrum of crystal materials, including 2D crystals, metal-organic frameworks (MOFs), defected crystals, and several sets of bulk crystals for comparison. Irregular and complex long-range order can be found extensively in the above-mentioned crystalline systems, while rarely exist in highly ordered traditional crystals. Detailed information about the benchmark is provided in Table~\ref{tab_datasets}.

The dataset \texttt{src\_bulk} in UnconvBench serves as the source of bulk crystals from which the dataset \texttt{defected} was derived. Although different targets are applied in these two datasets, they evaluate models in a sequential manner. The datasets \texttt{bulk\_s}, \texttt{bulk\_m}, and \texttt{bulk\_l} assess models using bulk crystals with varying sizes of crystal cells, addressing the prevalent weakness in many GNN-based machine learning models to learn global representations of large graphs. The dataset \texttt{bulk\_s} contains the smallest crystals, while the dataset \texttt{bulk\_l} contains the largest.

\begin{table*}[h]
\caption{Details of the 11 datasets in the benchmark, including target properties, number of samples, type of crystal samples and the source of data.}
\label{tab_datasets}
\begin{center}
\small
\resizebox{\textwidth}{!}{
\begin{tabular}{c|cccc}
\hline
\multicolumn{1}{c|}{\bf Datasets}  &\multicolumn{1}{c}{\bf Targets}  &\multicolumn{1}{c}{\bf Number of Sample}  &\multicolumn{1}{c}{\bf Material Category}   &\multicolumn{1}{c}{\bf Source of Data}                  
\\ \hline 
\texttt{2d\_e\_exf}& Exfoliation energy of 2D crystals ($\displaystyle eV/atom $)& 4,527& 2D crystal & \citep{gjerding2021c2db, haastrup2018c2db, choudhary2020jarvis}\\
\texttt{2d\_e\_tot} & Total formation energy of 2D crystals ($\displaystyle eV $)& 3,520& 2D crystal & \citep{zhou20192dmatpedia, choudhary2020jarvis}\\
\texttt{2d\_gap} & Band gap of 2D crystals ($\displaystyle eV $)& 3,520& 2D crystal & \citep{zhou20192dmatpedia, choudhary2020jarvis}\\
\texttt{co2\_adsp} & {CO$_2$} adsorption at 2.5 bar of MOFs ($wt\% $)& 13,765& MOF & \citep{bobbitt2016hmof, choudhary2020jarvis}\\
\texttt{qmof} & Formation energy of MOFs ($\displaystyle eV $)& 5,106& MOF & \citep{rosen2021qmof, choudhary2020jarvis}\\
\texttt{supercon} & Curie temperature (K)& 1,058& defected crystal & \citep{stanev2018supercon, choudhary2022supercon, choudhary2020jarvis}\\
\hline 
\texttt{defected}& Formation energy of defects ($\displaystyle eV/atom $)& 530& defected crystal & \citep{choudhary2022vacancy, choudhary2020jarvis}\\
\texttt{src\_bulk}& Formation energy of crystals ($\displaystyle eV/atom $)& 530& bulk crystal & \citep{choudhary2020jarvis, jain2013materialsproject}\\
\texttt{bulk\_s}& Formation energy of crystals ($\displaystyle eV/atom $)& 5,000& bulk crystal & \citep{dunn2020matbench, jain2013materialsproject}\\
\texttt{bulk\_m}& Formation energy of crystals ($\displaystyle eV/atom $)& 5,000& bulk crystal & \citep{dunn2020matbench, jain2013materialsproject}\\
\texttt{bulk\_l}& Formation energy of crystals ($\displaystyle eV/atom $)& 5,000& bulk crystal & \citep{dunn2020matbench, jain2013materialsproject}\\
\hline

\end{tabular}}
\end{center}
\end{table*}

\begin{figure}[t]
    \begin{center}
        \includegraphics[height=6.2cm]{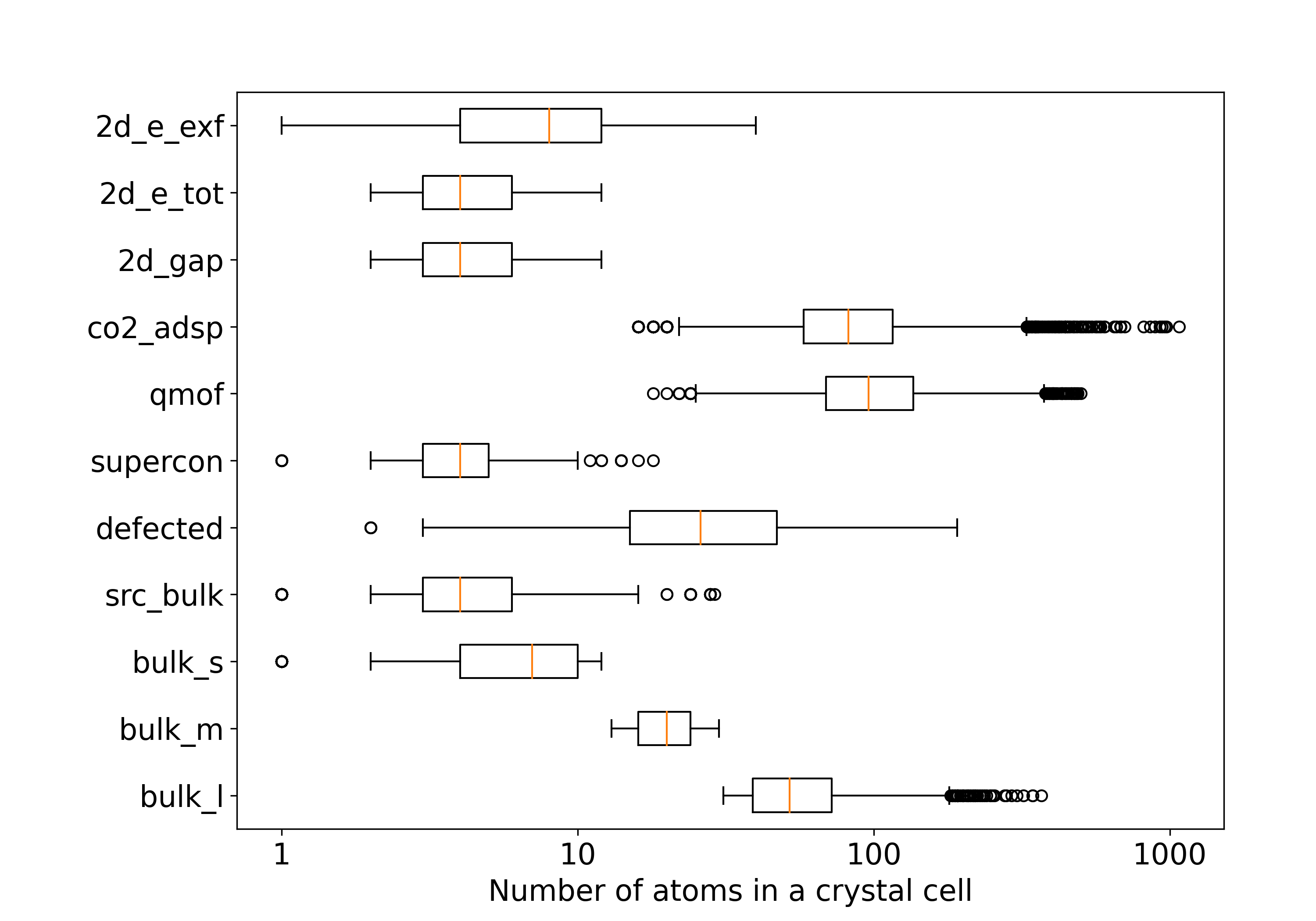}
    \end{center}
\caption{Statistics of crystal cell size in all datasets presented in box diagram. The MOF datasets \texttt{co2\_adsp} and \texttt{qmof} contain largest crystal cells, and the dataset \texttt{supercon} have smallest crystal cells. The crystal cells in \texttt{bulk\_s} exhibit a range of sizes from 1 to 12, whereas \texttt{bulk\_m} and \texttt{bulk\_l} contain crystals with sizes ranging from 13 to 30 and 31 to 368, respectively.}\label{fig_datasets}
\end{figure}

All datasets use crystal structures as raw input and a specific property as the prediction target. As illustrated in Figure~\ref{fig_datasets}, the largest crystal cell contains 500 atoms, while the smallest consists of only one. The average number of atoms in a single crystal cell ranges from 5 to 114. This variety allows the evaluation of models' performance on crystals of different sizes, stemming from the inclusion of various types of materials and deliberate selection. MOFs typically have larger repetitive units due to their porous structures, whereas bulk crystals have denser structures with smaller repetitive units. We divided a Materials Project \citep{jain2013materialsproject} dataset, \texttt{mp\_e\_form}, into three parts based on crystal cell size and sampled 5,000 crystals from each to create three datasets aimed at evaluating performance on variously sized crystal cells.

The target properties vary, ranging from formation energy for bulk crystals, exfoliation energy for low-dimensional crystals, to experimental properties such as Curie temperature for superconductors. More details of the datasets can be found in the appendices.

\subsection{Exploration of Model Structure}

Here, we mainly demonstrate the functionality of the two major components of the CrysToGraph model: the eTGC and the GwT by conducting an ablation study, and the proper structure to assemble the two parts by an enumeration. All validation losses are presented in terms of mean absolute error (MAE). The experiments are conducted on datasets \texttt{mp\_e\_form} and \texttt{log\_gvrh} from MatBench \citep{dunn2020matbench}. More training and optimization details can be found in the appendices.

\textbf{Functionality of eTGC and GwT}  The eTGC layers, as a message passing block, are designed to capture short-range dependencies and contribute a majority of the overall performance, while the GwT layers are primarily designed to model long-range interactions but also have a limited ability to capture short-range interactions. Although GwT layers effectively capture long-range interactions across crystal cells, its fully connected nature makes it less focused on the local environment around each node. 

\begin{figure}[t]
\begin{center}
%\framebox[4.0in]{$\;$}
\includegraphics[height=5cm]{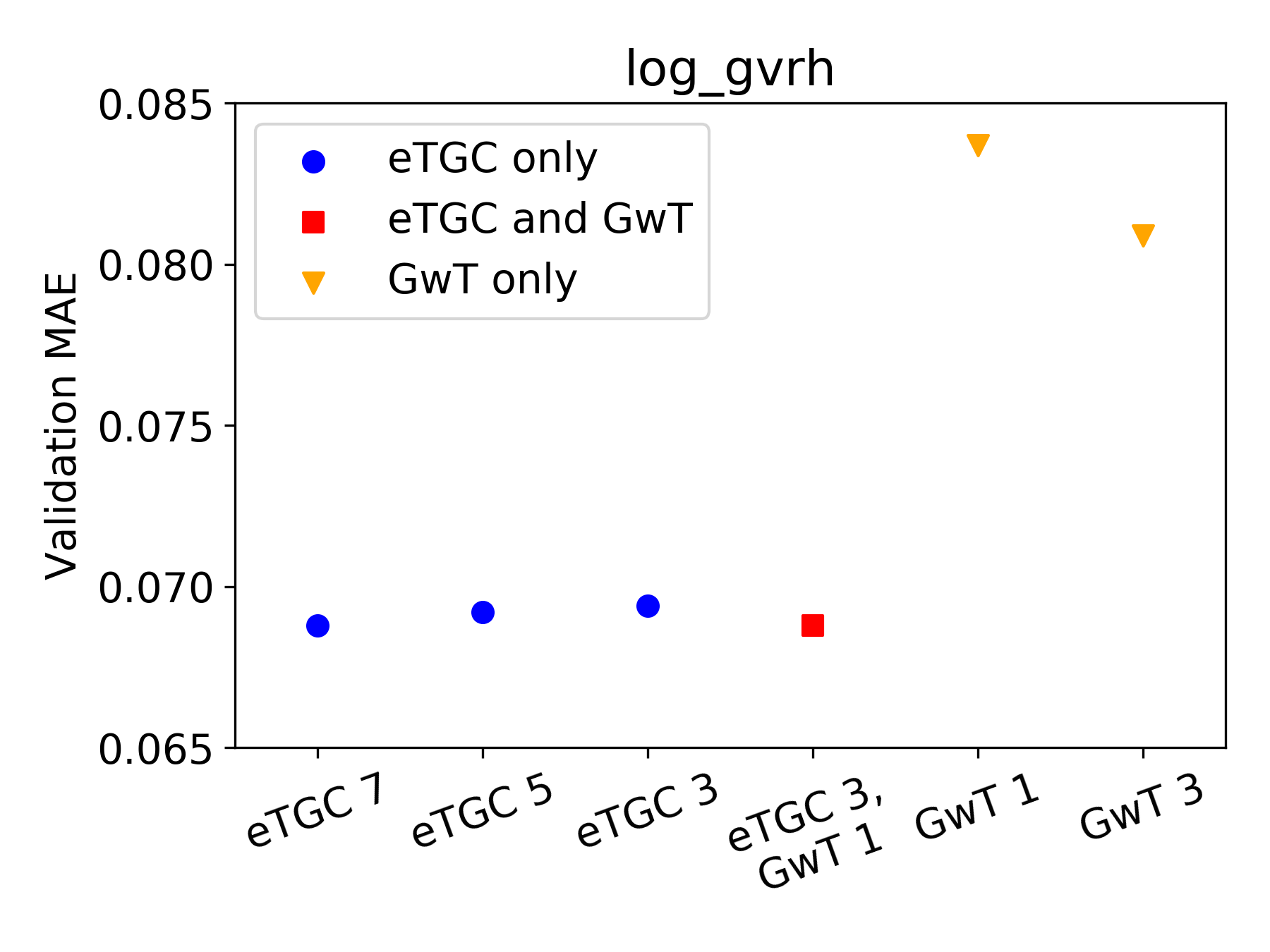}
\end{center}
\caption{Ablation studies on the functionality of eTGC and GwT conducted on \texttt{log\_gvrh} dataset. These models with different structures were train and validated on \texttt{log\_gvrh} in five-fold basis aligned with original the task. The eTGC 7 variant comprises 7 eTGC blocks without any GwT blocks, whereas the eTGC 3, GwT1 variant includes 3 eTGC blocks and 1 GwT block. Other data points are similarly documented.}\label{fig_gvrh}
\end{figure}

\begin{figure}[t]
\begin{center}
%\framebox[4.0in]{$\;$}
\includegraphics[height=5cm]{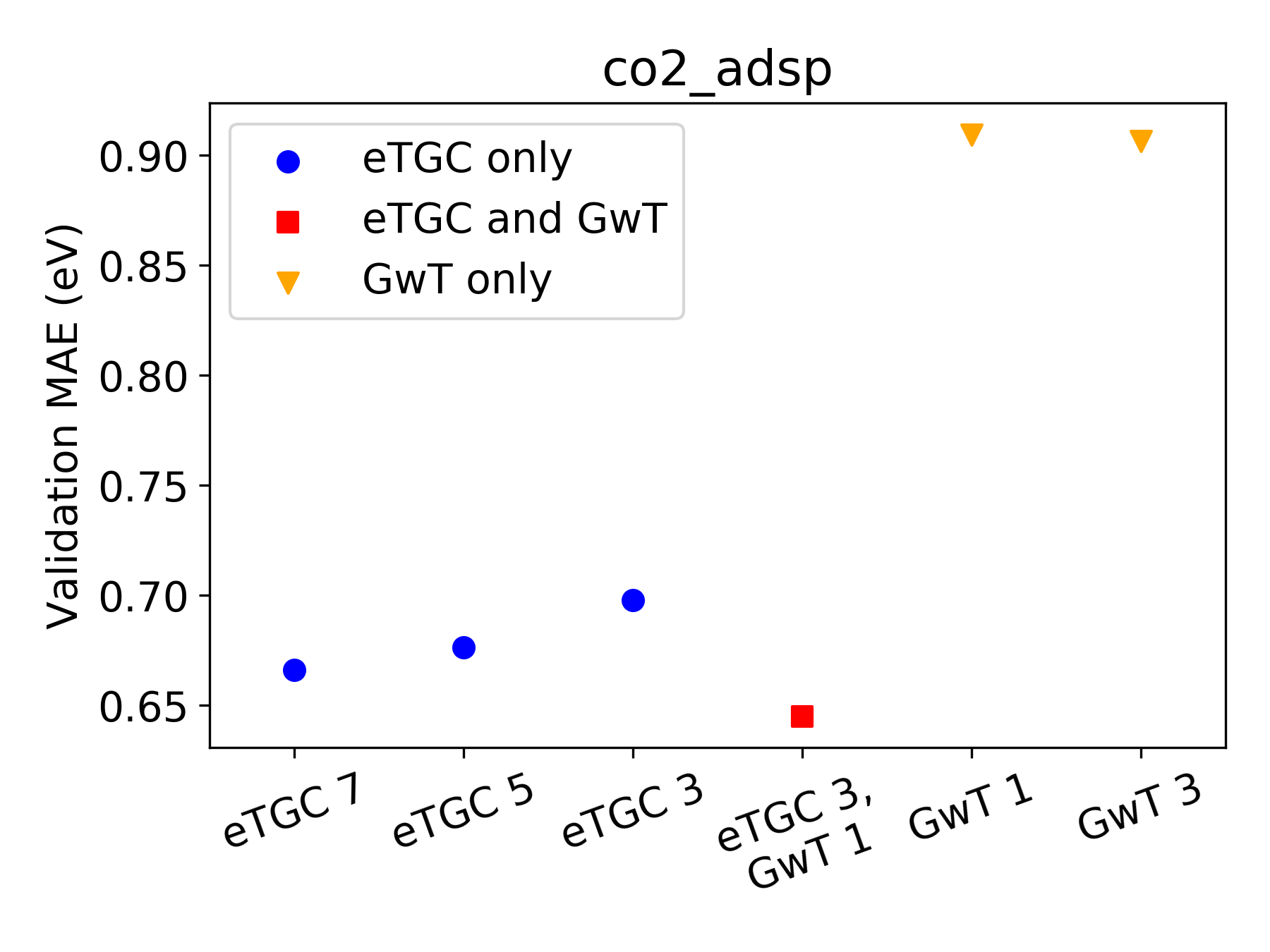}
\end{center}
\caption{Ablation studies on the functionality of eTGC and GwT conducted on \texttt{co2\_adsp} dataset. These models with different structures were train and validated on \texttt{co2\_adsp} in five-fold basis aligned with original the task. The notations are similar with Figure~\ref{fig_gvrh}.}\label{fig_co2}
\end{figure}

The functionalities of the two components are demonstrated on dataset \texttt{log\_gvrh}. As shown in Figure~\ref{fig_gvrh} and Figure~\ref{fig_co2}, the results were similar on the two datasets, indicating that: deeper networks generally lead to lower validation loss, which indicates better performance. However, either component on its own was hardly comparable to the merged eTGC-GwT merged models. 

Comparing the two figures in Figure~\ref{fig_gvrh} and Figure~\ref{fig_co2}, we can see a difference in performance on small crystal graphs and large crystal graphs. Notably, from the experiments on \texttt{log\_gvrh}, the eTGC blocks capture a major part of overall interactions. For small crystal graphs, the performance characteristics suggest that deeper eTGC blocks are sufficient for modeling long-range interactions within such crystal cells. However, as the sizes of the input crystal graphs increased, it became unrealistic to increase the depth without limit. In comparison, as shown in Figure~\ref{fig_co2}, when sizes of the crystal graphs increased, the effectiveness of the GwT structure at capturing long-range interactions became significant, indicating its crucial benefit to the graph network, even if the GwT structure has less peak performance on its own.

\begin{figure}[t]
\begin{center}
%\framebox[4.0in]{$\;$}
\includegraphics[height=5cm]{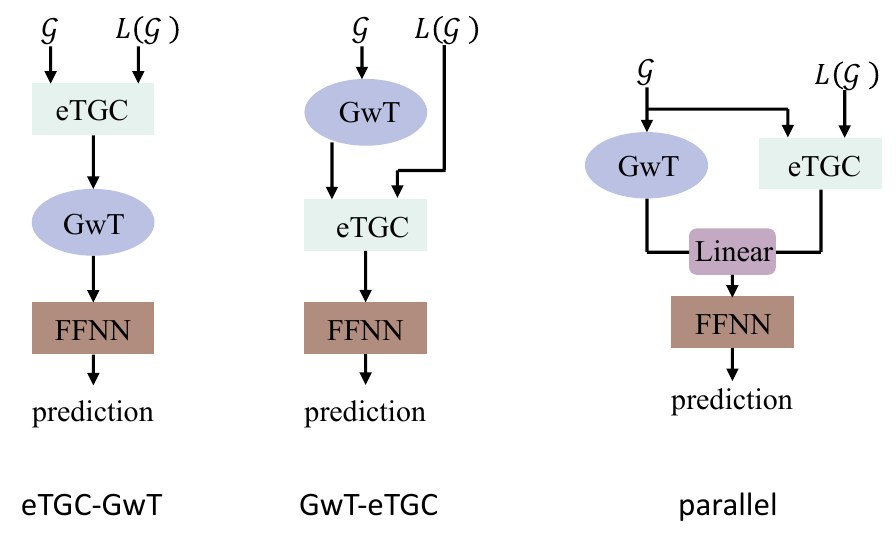}
\end{center}
\caption{Illustration of the three types of relationship between 3-blocked eTGC and GwT blocks: stacking in the order of eTGC-GwT, stacking in the order of GwT-eTGC and concatenating after parallel modelling.}\label{fig_merging}
\end{figure}

\textbf{Merging eTGC and GwT}  The relative positioning of the two major components impacted the successful modeling of crystals. As illustrated in Figure~\ref{fig_merging}, we investigated three hypotheses about the proper structure combination for modeling short-range and long-range chemical interactions: stacking in sequence with eTGC first, stacking in sequence with GwT first and running in parallel. We test these hypotheses with fixed depth of eTGC and GwT.

\begin{figure}[t]
\begin{center}
%\framebox[4.0in]{$\;$}
\includegraphics[height=5cm]{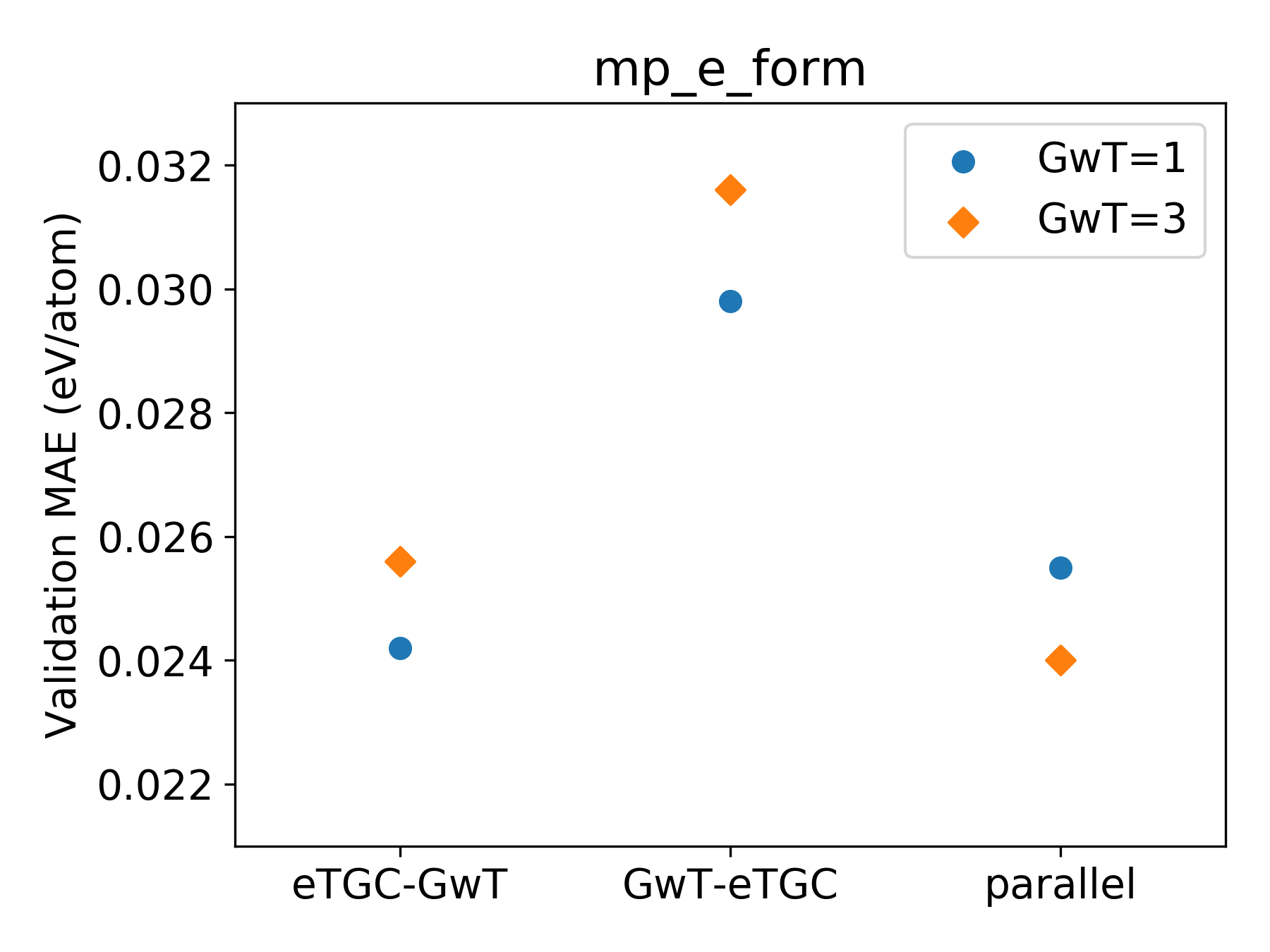}
\end{center}
\caption{Performances on a 10\% random subset of \texttt{mp\_e\_form} dataset with three types of relationship. The stacking structure in the order of 3 blocks of eTGC then 1 block of GwT performed similar as the parallel structure with 3 blocks of eTGC and 3 blocks of GwT. We adopted the former one as the final structure as it is simpler in structure and run faster.}\label{fig_stacking}
\end{figure}

The optimal depth of GwT depended on how the two components were assembled. With a parallel structure, deeper GwT blocks tend to yield better results. However, the parallel structure overall achieved worse peak performance than stacking structures. Of the various stacking structures, multiple eTGC layers followed by a single GwT layer outperformed other combinations. More specifically, in the experiments conducted on \texttt{mp\_e\_form} shown in Figure~\ref{fig_stacking}, we found that eTGC-GwT stacking structured models performed well when the GwT segment was shallow, but parallel structured models performed well when the GwT segment was deep. The validation loss of a parallel model with 3 layers of GwT was at the same level with a eTGC-GwT stacking model with 1 GwT layer, despite the latter being much simpler computationally. This result again proves that the GwT structure is effective at capturing long-range interaction and is an essential addition to the graph network model.

\subsection{Evaluation on Various Sized Cells and Defected Crystals  }

\begin{figure}[t]
\begin{center}
%\framebox[4.0in]{$\;$}
\includegraphics[height=5cm]{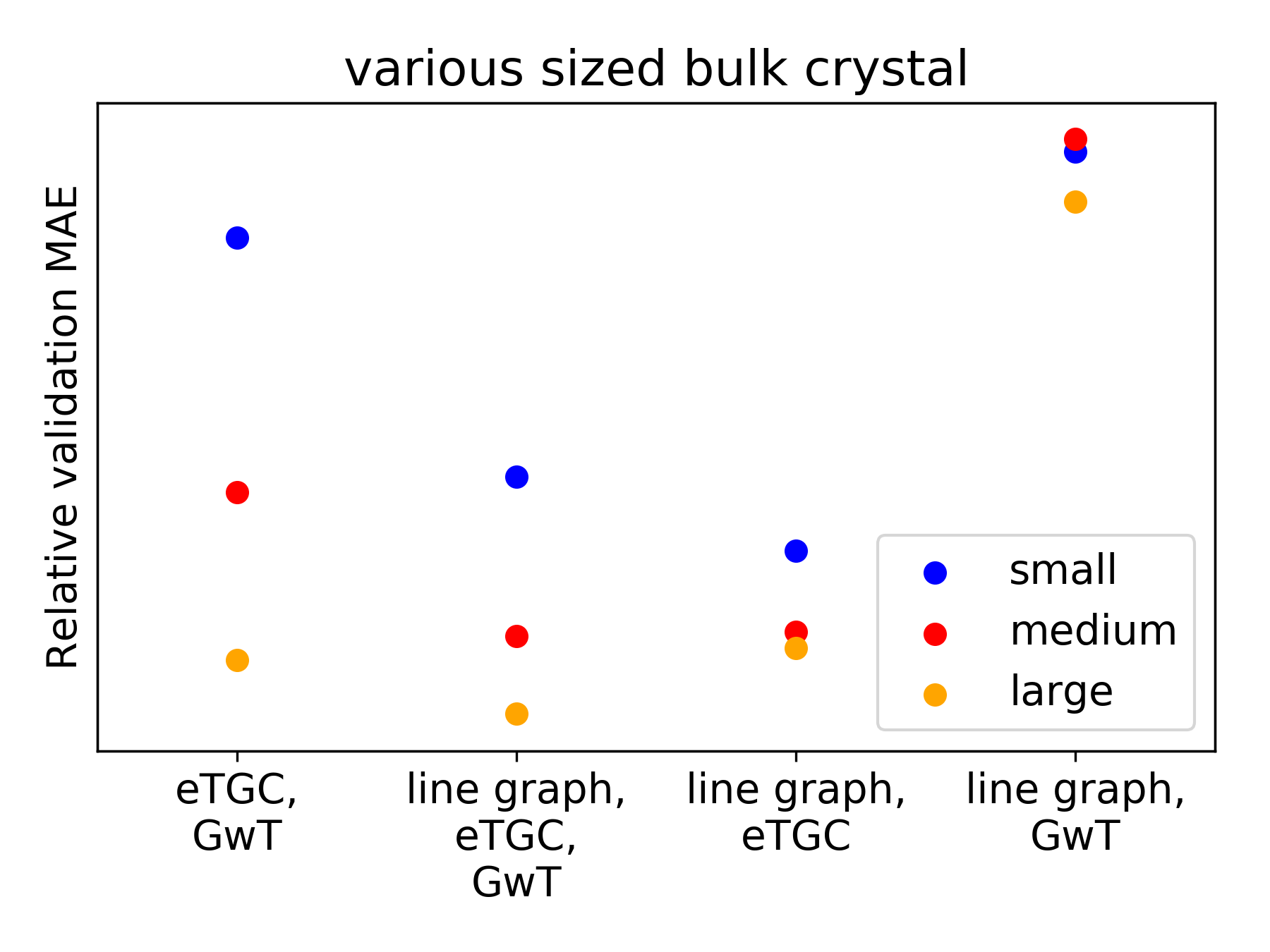}
\end{center}
\caption{Studies on the functionality of each component in this model: line graph, eTGC and GwT conducted on the curated datasets of various crystal cell size. The significance of GwT increases as the size of crystal cells increases. The column 2 is the full CrysToGraph model with line graphs, eTGC and GwT. In this section, the depth is 3 blocks of eTGC and 1 blockof GwT. The results here is shown in relative validation MAE loss due to the difference in the baseline of each dataset. This figure is demonstrated in a relative MAE as the baseline of the three tasks vaies.}\label{fig_size}
\end{figure}

We evaluated CrysToGraph and its variants, each with a critical component muted, on the curated datasets. These curated datasets include three datasets sampled from bulk crystals, sorted by the size of their repetitive units, as well as two datasets of defected and bulk counterparts.

In Figure~\ref{fig_size}, the model without graph-wise GwT layers performed best on the dataset with small cell-sized crystals but fell behind the full CrysToGraph model with GwT layers as the crystal cell size increased. This trend does not suggest that GwT is the primary component for extracting information from the graph, as the GwT-only model maintained the highest validation loss across all datasets. This trend aligns with the design philosophy of this work, where eTGC models the short-range interactions as the primary interactions within the crystal, and GwT models the long-range interactions as a secondary but significant aspect, particularly in crystals with large repetitive units where interactions and long-range order can be complex. These complex long-range orders are more common in certain types of crystal materials, such as MOFs and defected crystals, than in traditional crystals.

\begin{figure}[t]
\begin{center}
%\framebox[4.0in]{$\;$}
\includegraphics[height=5cm]{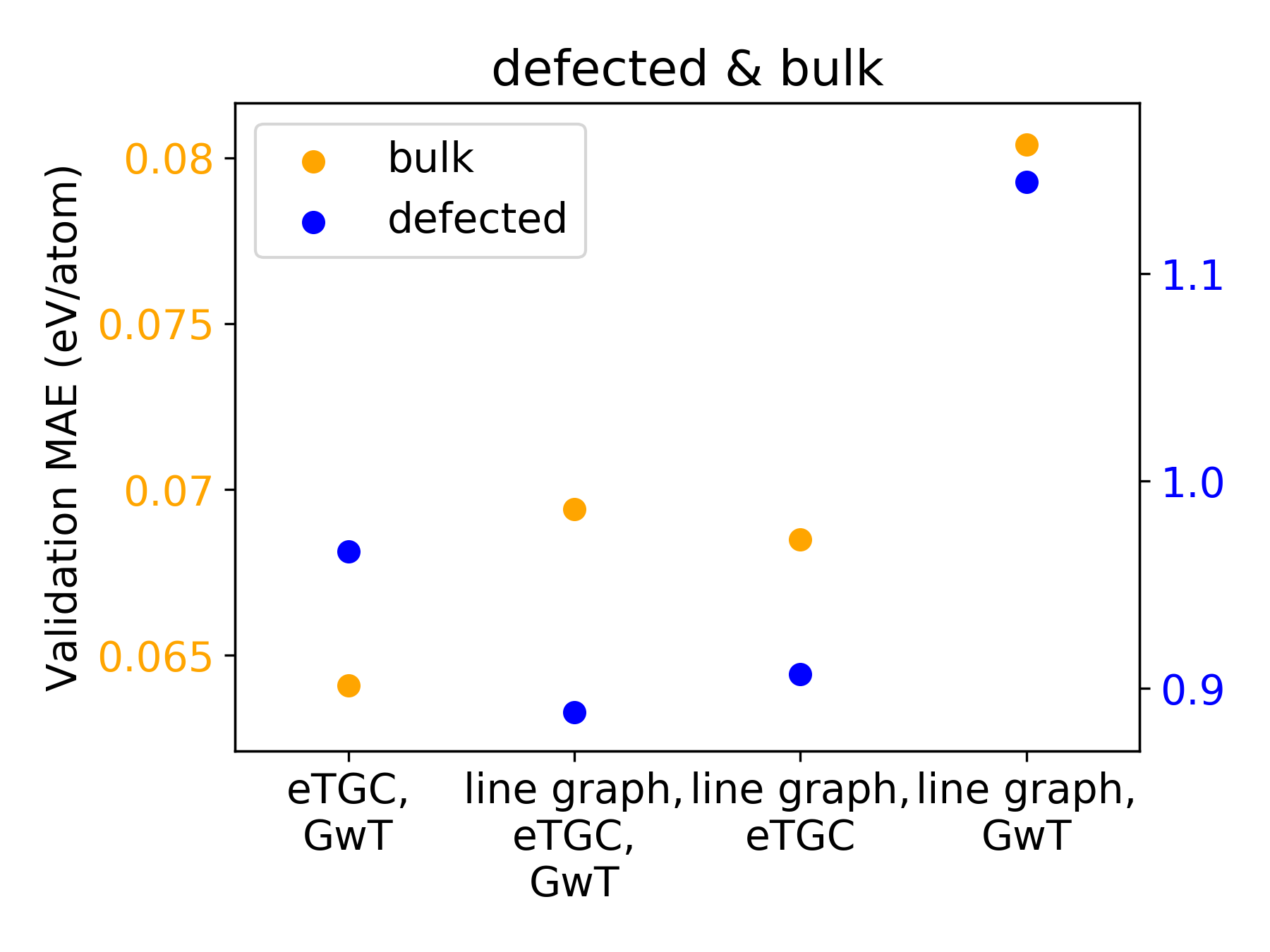}
\end{center}
\caption{Studies on the functionality of each component in this model: line graph, eTGC and GwT conducted on the curated datasets of bulk and defected crystals. The GwT layers and line graphs are more important for the successful modelling of defected crystals than the bulk crystals. The column 2 is the full CrysToGraph model with line graphs, eTGC and GwT. The structure is 3 blocks of eTGC and 1 block of GwT here, aligned with the studies on various sized bulk crystal.}\label{fig_vacancy}
\end{figure}

In Figure~\ref{fig_vacancy}, the performance trend of models on defected and bulk crystals further supports our design rationale. The samples in the \texttt{defected} dataset were derived from the source crystals in \texttt{src\_bulk}, with random defects such as insertion, vacancy, and substitution introduced at probabilities consistent with literatures \citep{choudhary2022vacancy}. The model without line graphs achieved the lowest error on bulk crystals, whereas the full CrysToGraph model performed best on defected crystals. We attribute this result to the line graphs, which convey structural information that aids the full model, particularly the GwT layers, in modeling the complex interactions and long-range order in defected crystals.

\subsection{Evaluation on Multiple Types of Materials}

The benchmark in Section 4.1 provides a comprehensive evaluation of predictive models on certain types of crystal materials. We present the predictive performance of CrysToGraph on this benchmark, comparing it with several baseline models in Table~\ref{tab_unconvbench}. All results are reported using mean absolute error (MAE) as the metric, and all tasks were trained using 5-fold cross-validation. Current state-of-the-art results are shown in \textbf{bold}, while second-place results are indicated with \underline{underlined}.

CrysToGraph achieved state-of-the-art results in all six tasks. Notably, there is a significant performance gap between CrysToGraph and the second-place model on the \texttt{qmof} dataset, which includes very large crystal cells. CrysToGraph, as the only model in the benchmark featuring a graph-wise attention layer, likely benefits from the GwT layers in effectively modeling the MOFs. CrysToGraph demonstrated the effectiveness of its explicit capture of long-range interactions by outperforming other models that less explicitly model these interactions. A comparison of the architecture of the models on the benchmark can be found in the appendices.
\begin{table*}[h]
\caption{Results on the general predictive tasks in the  UnconvBench. All results are shown in the mean of validation MAE in five-fold cross validation. The details of datasets can be found in Table~\ref{tab_datasets}.}
\label{tab_unconvbench}
\small
\begin{center}
\resizebox{\textwidth}{!}{
\begin{tabular}{c|ccccccc}
\hline
\multicolumn{1}{c|}{\bf Datasets}  &\multicolumn{1}{c}{\bf CrysToGraph}  &\multicolumn{1}{c}{\bf coGN\citep{ruff2023cogn}}  &\multicolumn{1}{c}{\bf coNGN\citep{ruff2023cogn}}  &\multicolumn{1}{c}{\bf ALIGNN\citep{choudhary2021alignn}}  &\multicolumn{1}{c}{\bf CGCNN\citep{xie2018cgcnn}} &\multicolumn{1}{c}{\bf MODNet\citep{de2021modnet, de2021modnet2}}  &\multicolumn{1}{c}{\bf Dummy}  
\\ \hline 
\texttt{2d\_e\_exf} & \bf{0.0500}& \underline{0.0510}& 0.0530& 0.0580& 0.0710& 0.0665& 0.1195\\
\texttt{2d\_e\_tot} & \bf{0.3623}& 0.5214& 0.4497& \underline{0.3705}& 1.2941& 1.5267& 13.0904\\
\texttt{2d\_gap} & \bf{0.0986}& 0.1168& 0.1432& \underline{0.1048}& 0.1499& 0.1612& 0.5631\\
\texttt{qmof} & \bf{121.9662}& 218.9272& 229.1948& \underline{217.2508}& 231.1887& 299.5225& 331.6826\\
\texttt{supercon} & \bf{2.6422}& 2.8955& 2.9167& \underline{2.7372}& 2.9316& 3.2210& 3.4705\\
\texttt{defected}& \bf{0.8885}& 1.0615& 1.0441& 0.9842& 1.1321& \underline{0.9168}& 1.8121\\
\hline

\end{tabular}}
\end{center}
\end{table*}

\subsection{Evaluation on General Crystal Benchmark}

\begin{table*}[h]
\caption{Results on traditional crystal benchmark MatBench, among which \texttt{mp\_is\_metal} is the only classification task. The training and validation was conducted using the MatBench \citep{dunn2020matbench} API. All results are shown in the mean of validation MAE in five-fold cross validation, aligned with the MatBench benchmark. The details of datasets can be found in the appendecies.}
\label{tab_matbench}
\small
\begin{center}
\resizebox{\textwidth}{!}{
\begin{tabular}{c|ccccccc}
\hline
\multicolumn{1}{c|}{\bf Datasets}  &\multicolumn{1}{c}{\bf CrysToGraph}  &\multicolumn{1}{c}{\bf coGN\citep{ruff2023cogn, dunn2020matbench}}  &\multicolumn{1}{c}{\bf coNGN\citep{ruff2023cogn, dunn2020matbench}}  &\multicolumn{1}{c}{\bf ALIGNN\citep{choudhary2021alignn, dunn2020matbench}}  &\multicolumn{1}{c}{\bf CGCNN\citep{xie2018cgcnn, dunn2020matbench}} &\multicolumn{1}{c}{\bf Matformer\citep{yan2022matformer, ruff2023cogn}} &  
\bf Dummy\citep{dunn2020matbench}\\ \hline 
\texttt{dielectric} & \bf{0.3084}& \underline{0.3088}& 0.3142 & 0.3449 & 0.5988 & 0.6340 &0.8088 \\
\texttt{jdft2d} & \bf32.3720  & \underline{37.1652}& 36.1698 & 43.4244 & 49.2440 & 42.8270 &67.2851\\
\texttt{log\_gvrh} & \underline{0.0686}& 0.0698 & \bf0.0670 & 0.0715 & 0.0895 & 0.0770 &0.2931\\
\texttt{log\_kvrh} & \underline{0.0519}& 0.0535& \bf{0.0491} & 0.0568 & 0.0712 & 0.0630 &0.2897\\
\texttt{mp\_e\_form} & \bf{0.0168}& \underline{0.0170}& 0.0178& 0.0215 & 0.0337 & 0.0212 &1.0059\\
\texttt{mp\_gap} & \bf{0.1522}& \underline{0.1559}& 0.1697& 0.1861 & 0.2972 & 0.1878 &1.3272\\
\texttt{\underline{mp\_is\_metal}}& \underline{0.9146}  & 0.9124 & 0.9089 & 0.9128 & \bf0.9520 & 0.9060 &0.5012\\
\texttt{phonons} & \bf{28.3990}  & 29.7117 & \underline{28.8874}& 29.5385 & 57.7635 & 42.5260 &323.9822\\
\hline

\end{tabular}}
\end{center}
\end{table*}

We further demonstrated the overall effectiveness of CrysToGraph and compared our model with others on various benchmarks, as shown in Table~\ref{tab_matbench}. Among the eight tasks, seven were regression tasks, while \texttt{mp\_is\_metal} was a classification task. The results for the regression tasks are presented using mean absolute error (MAE), while the result for the classification task is presented using the area under the receiver operating characteristic curve (AUC). In alignment with the evaluation and presentation of results on UnconvBench, all tasks were trained with 5-fold cross-validation; current state-of-the-art results are shown in \textbf{bold} font and second place is shown \underline{underlined}.

In this comparison, all other models consider only the local environments, with Matformer \citep{yan2022matformer} being the transformer-based model among them. Of the eight tasks, CrysToGraph achieved state-of-the-art results in five and secured second place in the remaining three tasks. On both large datasets of over 100,000 samples (\texttt{mp\_gap} and \texttt{mp\_e\_form}) and small datasets of hundreds of samples (\texttt{phonons} and \texttt{jdft\_2d}), CrysToGraph outperformed other models. It is worth noting that the crystals in the \texttt{jdft2d} dataset are also 2D crystals, on which the performance of CrysToGraph was significantly better than others. CrysToGraph, designed to separately capture short-range and long-range interactions, has proven competitive in modeling traditional crystals as well.

%% file: main/discussion.tex
\section{Discussion}

\subsection{Correlation of Predictions and Ground Truth}\label{sec_discussion_1}

\begin{figure}[h]
\begin{center}
%\framebox[4.0in]{$\;$}
\includegraphics[height=6cm]{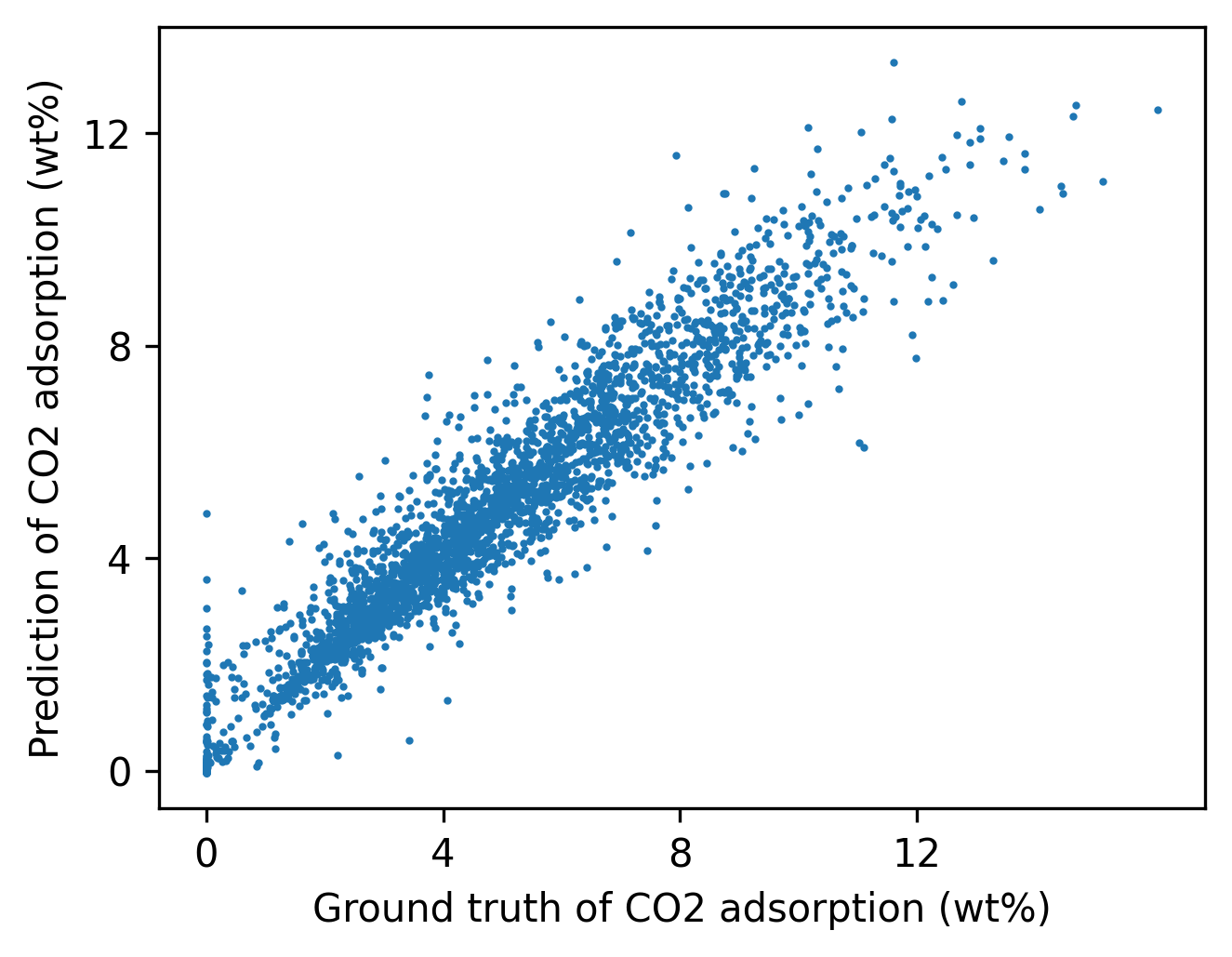}
\end{center}
\caption{Scatter diagram of prediction and ground truth in task \texttt{co2\_adsorption}. x axis represents the label as the ground truth, y axis represents the predictions.} \label{fig_corr_co2}
\end{figure}

We scattered the prediction of CrysToGraph model over the ground thruth in the task \texttt{co2\_adsp} in Figure~\ref{fig_corr_co2}, where high accuracy of the CrysToGraph model's predictions is illustrated, with all 5-fold test results plotted. The $r^2$ score for this task is calculated to be 0.8946. The correlation in other tasks might be stronger or weaker depends on the tasks and dataset.

\section{Conclusion}

In this paper, we introduced CrysToGraph, a geometric graph neural network designed to capture both short-range and long-range interactions in crystal materials. Our model utilizes transformer-based message-passing blocks (eTGC) and graph-wise transformers (GwT) to effectively capture these interactions. Through our evaluation on the MatBench and UnconvBench benchmarks, we demonstrated that our model outperformed existing approaches in 11 out of 14 tasks, establishing new state-of-the-art results.

One key finding of our study is the distinct roles played by the eTGC blocks and GwT layers. We observed that the eTGC blocks primarily capture short-range interactions, while the GwT layers are responsible for capturing long-range interactions. This understanding of the model's components and their assembly provides valuable insights into the underlying mechanisms of crystal materials.

To facilitate further research in the field, we introduced the UnconvBench benchmark, a benchmark to comprehensively evaluate predictive models on multiple types of crystal materials. This benchmark encompasses various types of unconventional crystals, including MOFs, 2D materials, defected crystals, and a curated selection of bulk crystals. By incorporating these diverse materials, UnconvBench enables a thorough evaluation of predictive models across a wide range of scenarios.

While our model achieved promising results on both benchmarks, we acknowledge that there is still room for improvement in performance. We did not pursue further optimization in this study, as the current results already demonstrate the significance of explicitly capturing short-range and long-range interactions in crystal graphs, as well as in graphs more broadly. However, we encourage readers to explore the appendices for detailed information on the model architecture, training procedures, and additional experimental results.

Looking ahead, we believe that CrysToGraph can be applied to real-world molecular dynamic simulations in crystalline systems, offering valuable insights into their behavior and properties. Furthermore, we envision the potential extension of our model to other chemical systems, opening up new avenues for research in the broader field of materials science.

In summary, our work presents CrysToGraph as an effective graph neural network for modeling crystal materials, showcasing its superior performance on established benchmarks and introducing a new benchmark for novel crystal materials. Our findings set up new state-of-the-art results on benchmarks and provide a new solution for the virtual screening 
and other downstream research in the field of crystal materials. We encourage readers to delve into the appendices for a comprehensive understanding of our methodology and results.

%% file: main/back.tex
\section{Declarations}

\subsection{Impact Statement}

This paper presents work whose goal is to advance the field of Machine Learning. There are many potential impact of our work in chemistry, materials science and machine learning, none which we feel must be specifically highlighted here.

\subsection{Conflict of interest}

The author of this work has not conflict of interest.

\subsection{Code Availability}

All data and code generated or analyzed during the current study are open-sourced. Codes about CrysToGraph model are available at: https://github.com/howardwang1997/CrysToGraph; codes about UnconvBench are available at: https://github.com/howardwang1997/unconvbench.

\subsection{Acknowledgement}

We would like to express our sincere gratitude to Prof. Yoshua Bengio, Prof. Jian Tang, Mr. Zhaocheng Zhu, Mr. Chuanrui Wang, Mr. Huiyu Cai from Mila, Montr\'eal, Canada for constructive discussion and advice.

\begin{appendices}
\section{Hyperparameters, training details and model structures}
\subsection{Masked Atom Pretraining}\label{secA1}

\begin{figure}[h]
\begin{center}
%\framebox[4.0in]{$\;$}
\includegraphics[height=6cm]{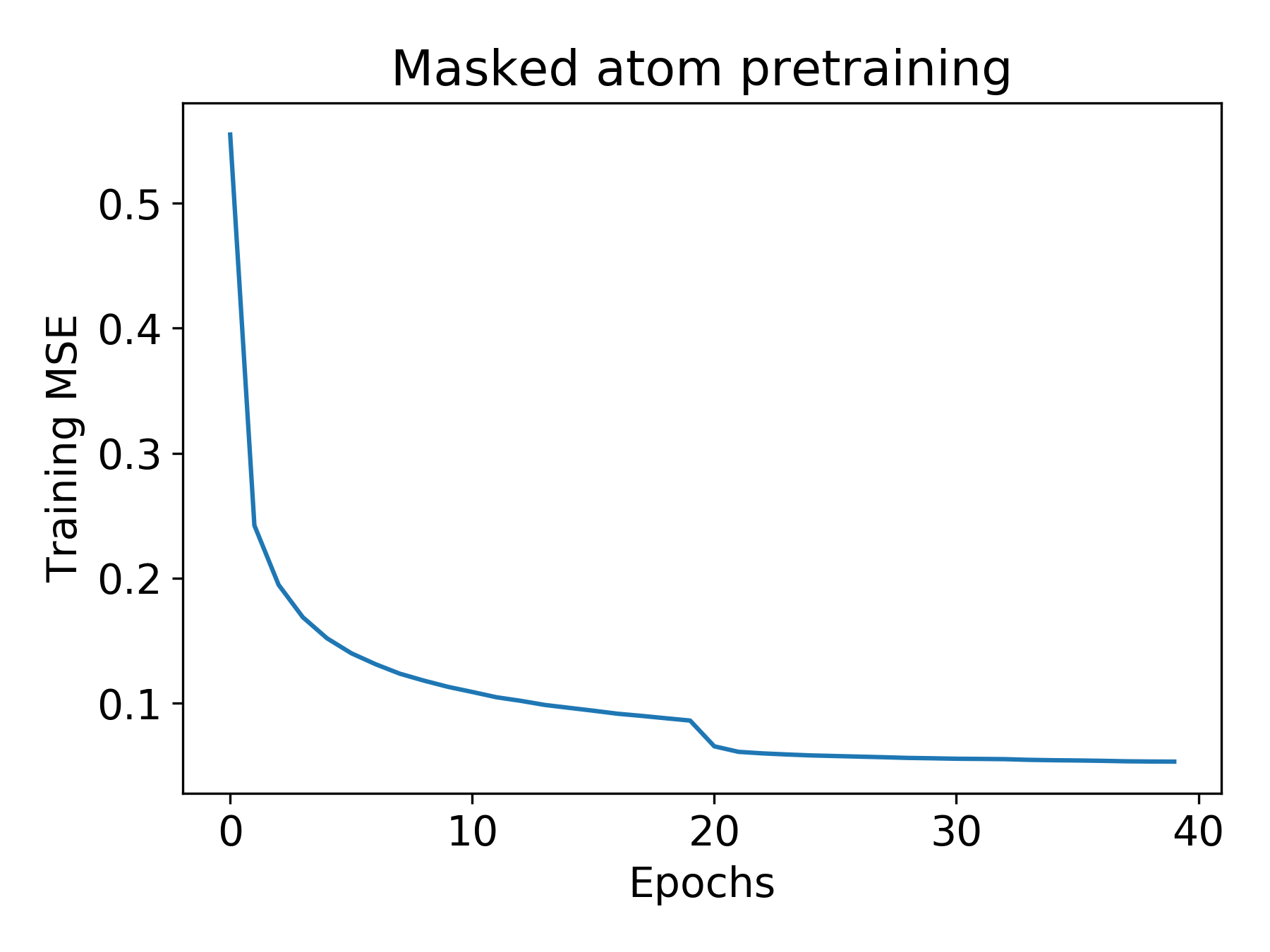}
\end{center}
\caption{The loss curve in atom masked pretraining. The learning rate decrease by 10 at epoch 20.} \label{figA_arp}
\end{figure}

In the pretraining phase for atom representations, we introduced a masked atom prediction task. In this task, a specified percentage of atoms within each graph are masked. Specifically, 15\% of the atoms in each graph are subjected to masking operations. Among these, 80\% are substituted with a designated mask token, while 10\% are replaced with randomly selected tokens, and the remaining 10\% are left unchanged. In instances where the number of nodes in a graph is insufficient to maintain the masking rate below 15\%, the crystal structure is expanded in all three dimensions. With graphs constructed as aforementioned, we trained a CGCNN model on these constructed graphs to predict the types of masked atoms. The loss curve is in Figure~\ref{figA_arp}.

Following the pretraining of atom embeddings, we concatenated the machine-learnt embeddings with the manually curated CGCNN atom embeddings.

\subsection{Layers of eTGC}\label{secA2}

\begin{figure}[h]
\begin{center}
%\framebox[4.0in]{$\;$}
\includegraphics[height=6cm]{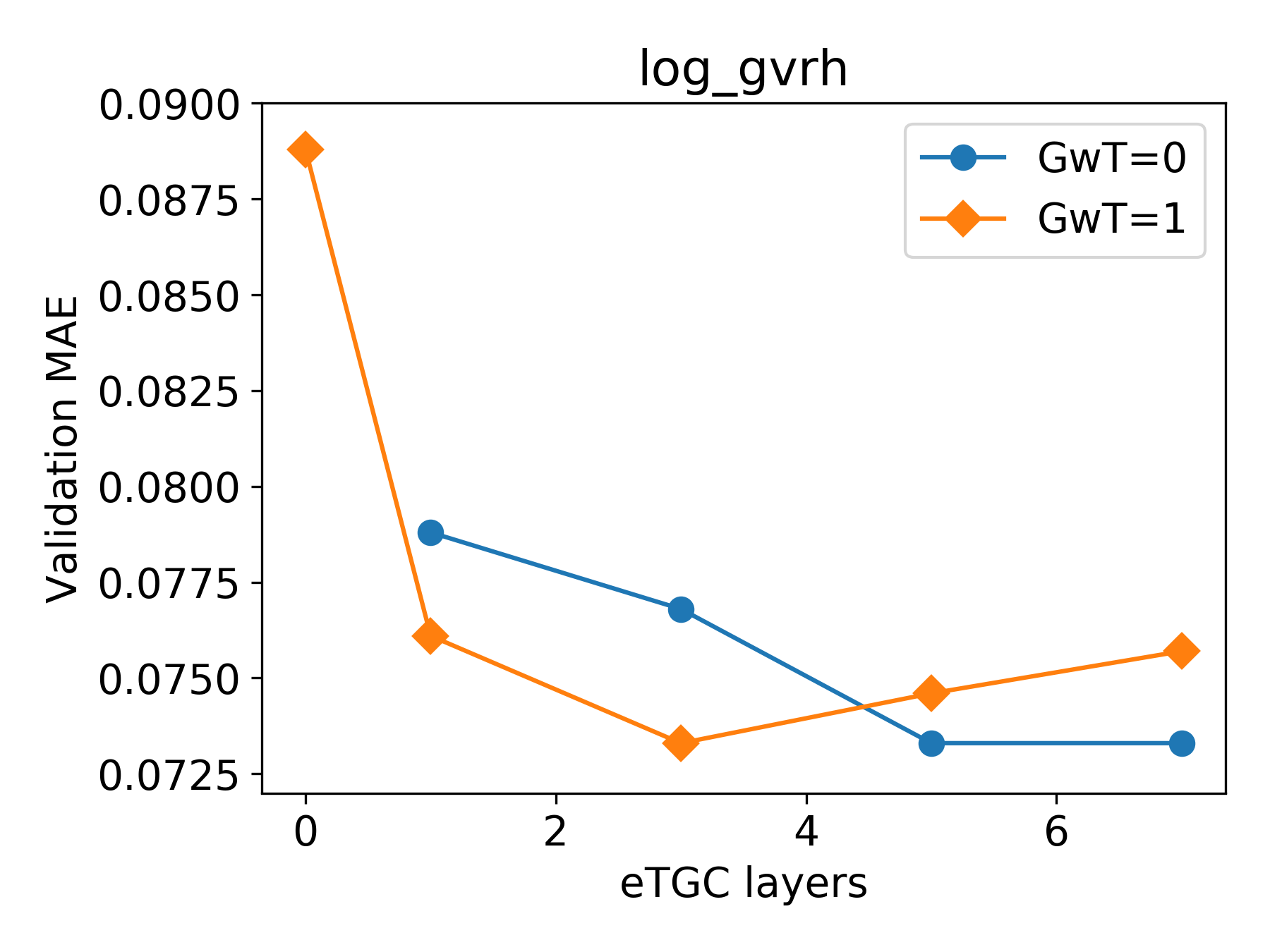}
\end{center}
\caption{Various depth of eTGC, trained on \texttt{log\_gvrh} dataset.} \label{figA_etgc_gvrh}
\end{figure}

\begin{figure}[h]
\begin{center}
%\framebox[4.0in]{$\;$}
\includegraphics[height=6cm]{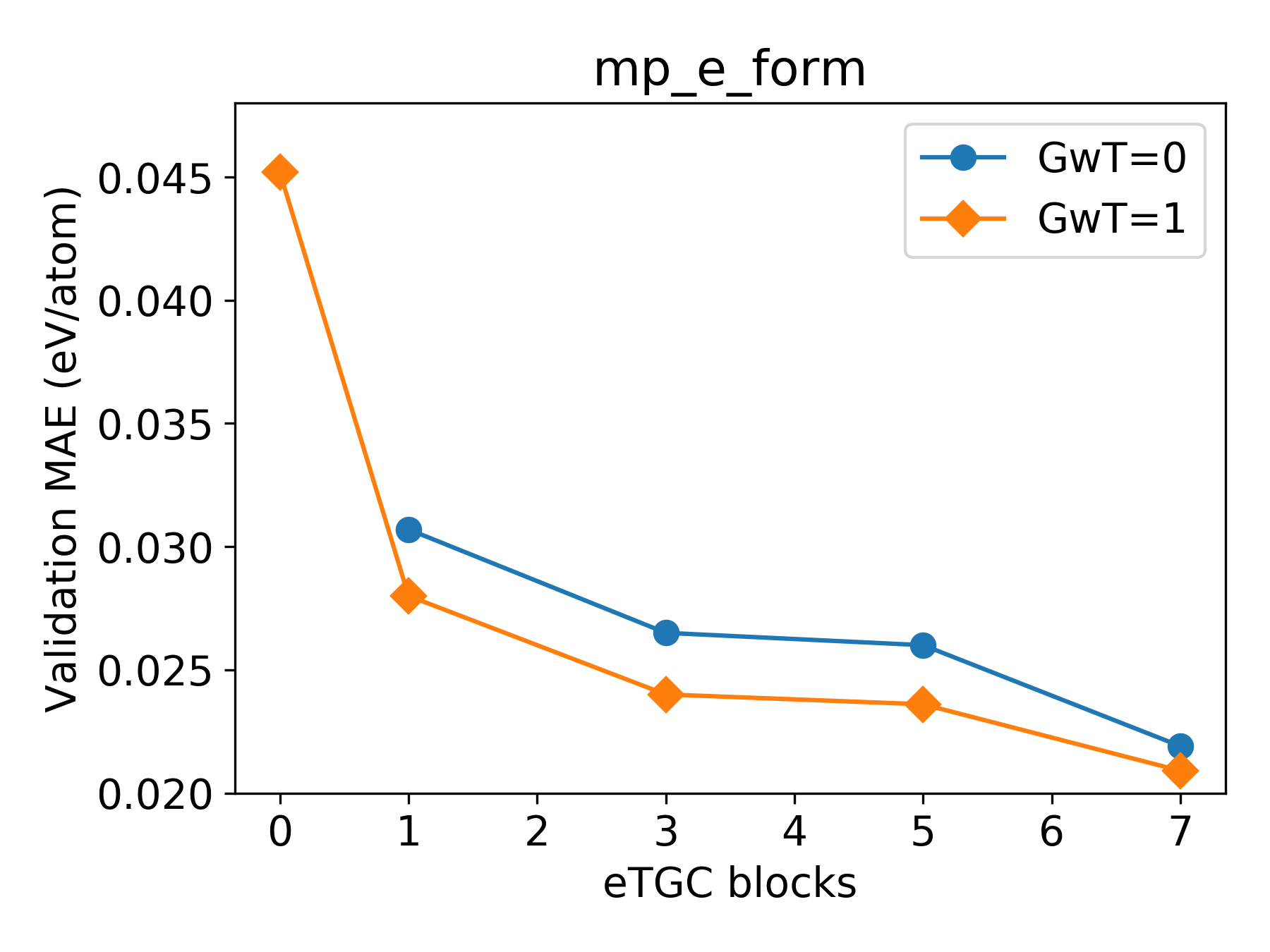}
\end{center}
\caption{Various depth of eTGC, trained on \texttt{mp\_e\_form} dataset.} \label{figA_etgc_eform}
\end{figure}

Figure~\ref{figA_etgc_gvrh} and Figure~\ref{figA_etgc_eform} demonstrates that deeper eTGC usually leads to lower validation loss, however, there are exceptions when stacking too deep eTGC wit GwT. In general, the ideal depth of eTGC is around 3 to 7, when the specific depth depends on dataset.

\subsection{Layers of GwT}\label{secA3}

\begin{figure}[h]
\begin{center}
%\framebox[4.0in]{$\;$}
\includegraphics[height=6cm]{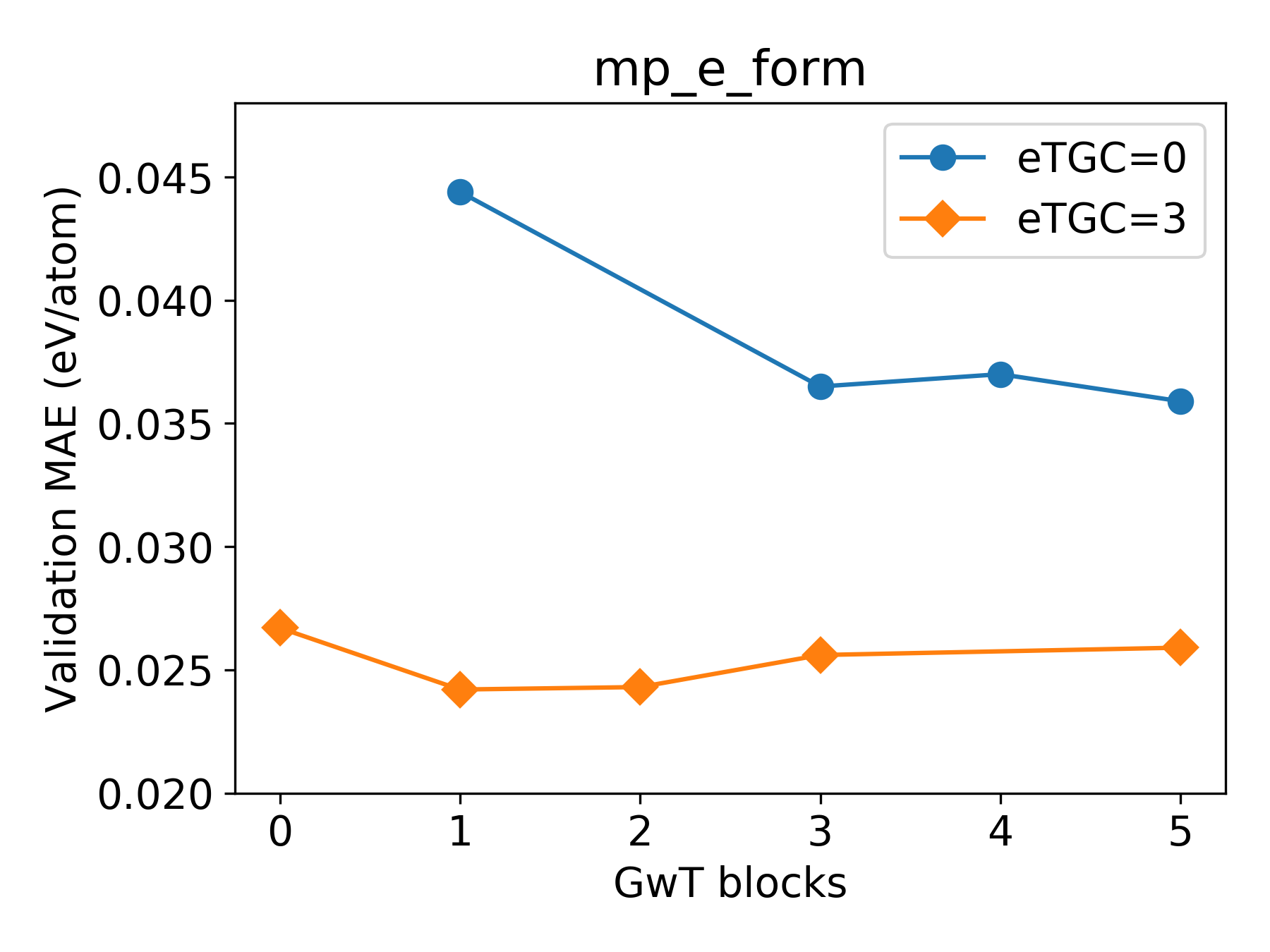}
\end{center}
\caption{Various depth of GwT, trained on \texttt{mp\_e\_form} dataset.} \label{figA_gwt}
\end{figure}

From Figure~\ref{figA_gwt}, we confirm that the GwT contributes the minor part in the overall network. Given eTGC exists, the ideal depth of GwT is 1 layer.

\subsection{Parallel eTGC and GwT}\label{secA4}

\begin{figure}[h]
\begin{center}
%\framebox[4.0in]{$\;$}
\includegraphics[height=6cm]{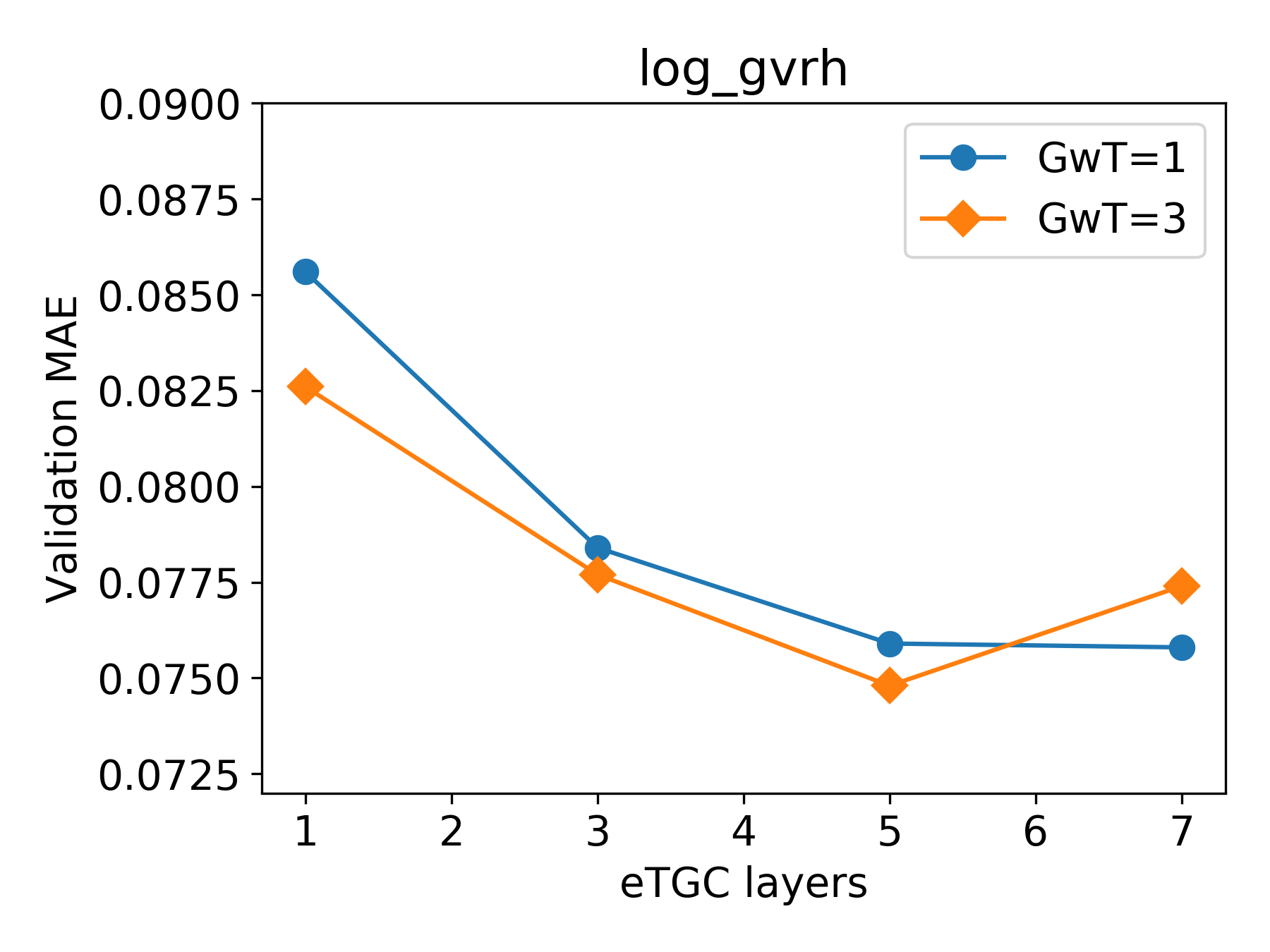}
\end{center}
\caption{Various depth of eTGC with fixed depth of GwT, in parallel structures, trained on \texttt{log\_gvrh} dataset.} \label{figA_parallel}
\end{figure}

This Figure~\ref{figA_parallel} demonstrates the behavior of CrysToGraph model when eTGC and GwT blocks cooperate in a parallel structure. Deeper GwT performs generally better than shallower GwT, except when eTGC is deep enough.

\subsection{Layers of FFNN}\label{secA5}

\begin{figure}[h]
\begin{center}
%\framebox[4.0in]{$\;$}
\includegraphics[height=6cm]{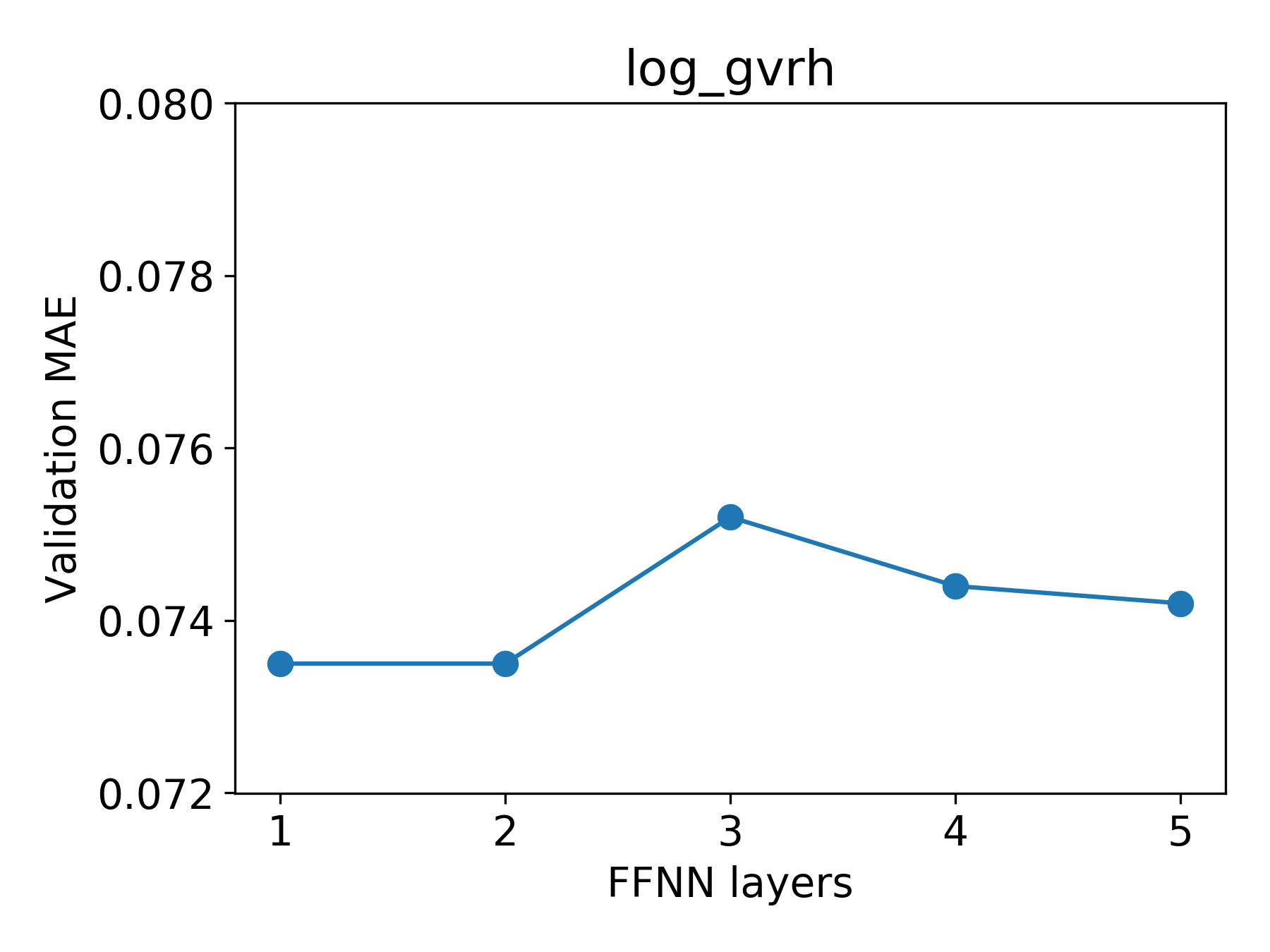}
\end{center}
\caption{Various depth of FFNN, trained on \texttt{log\_gvrh} dataset.} \label{figA_fc}
\end{figure}

We can see in Figure~\ref{figA_fc}, deeper FFNN impedes the overall performance. The ideal depth of FFNN is 1 or 2 layer.

\subsection{Learning Rate}

\begin{figure}[h]
\begin{center}
%\framebox[4.0in]{$\;$}
\includegraphics[height=6cm]{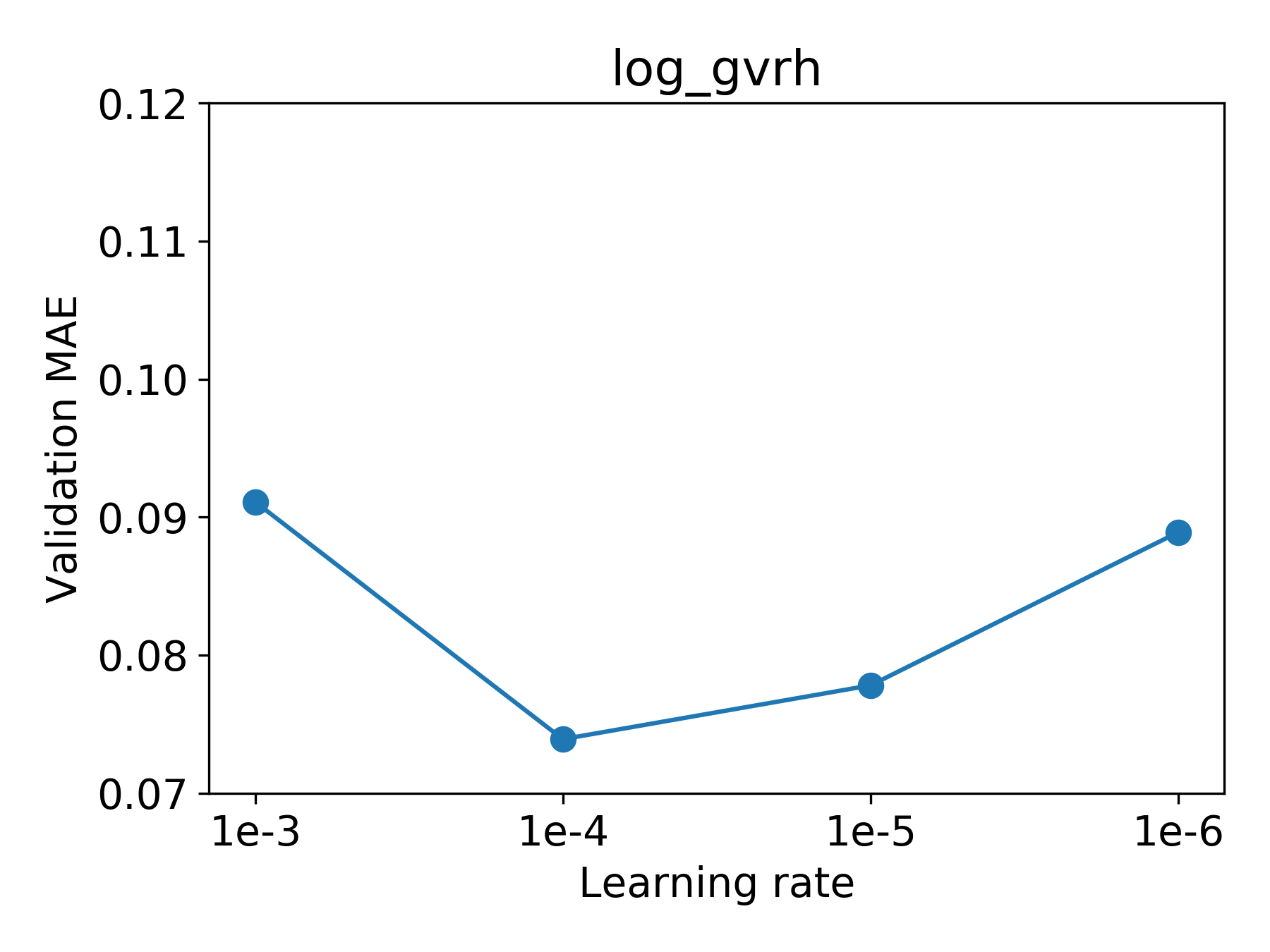}
\end{center}
\caption{Same model trained in various learning rate.} \label{figA_lr}
\end{figure}

Shown in Figure~\ref{figA_lr}, the optimal learning rate is 1e$^{-4}$, which is generalized to other experiments in this work.

\subsection{Weight Decay}\label{secA6}

\begin{figure}[h]
\begin{center}
%\framebox[4.0in]{$\;$}
\includegraphics[height=6cm]{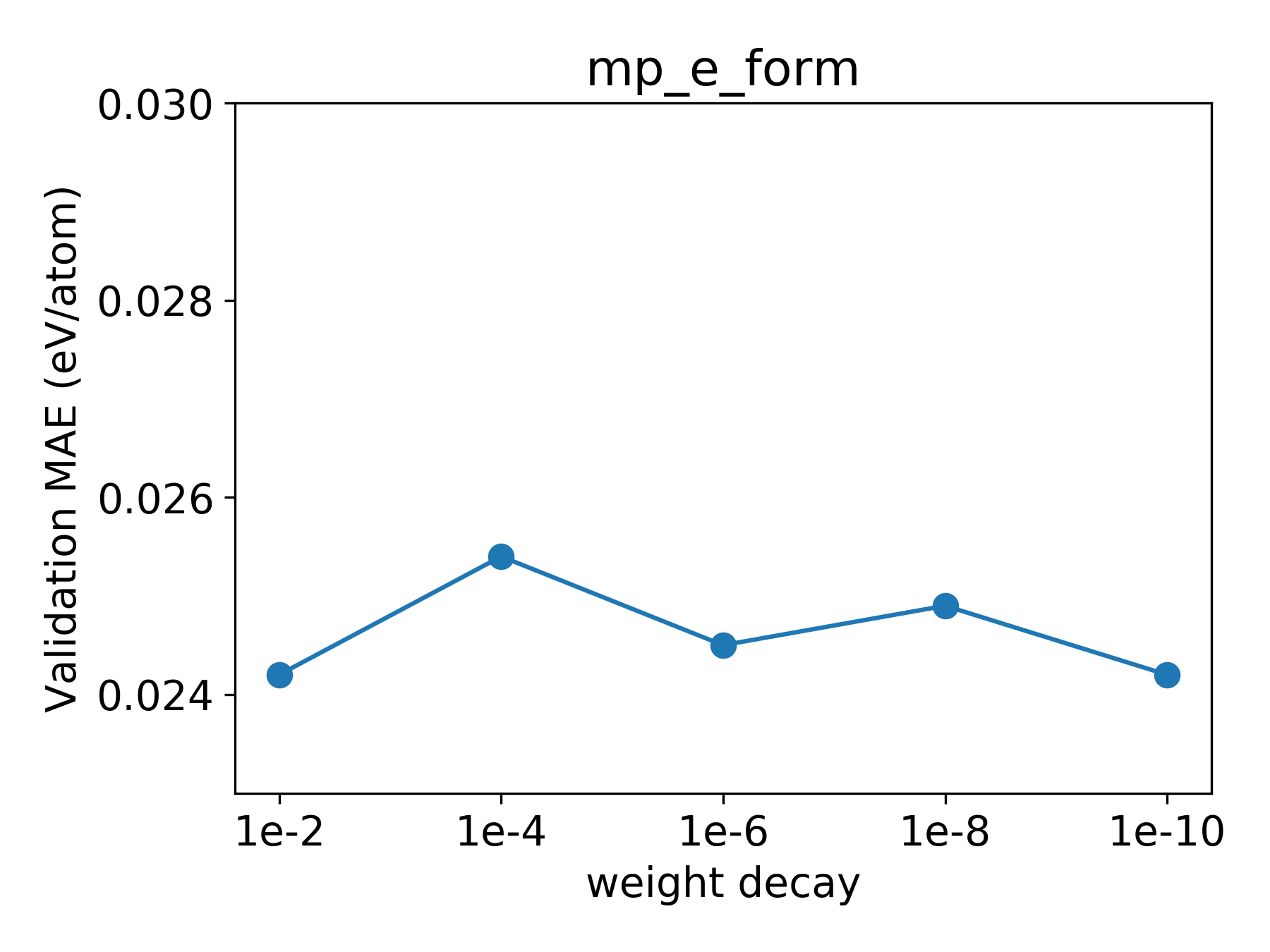}
\end{center}
\caption{Models trained with various weigh decay penalty.} \label{figA_decay}
\end{figure}

The weight decay can be regarded as a derivative of L2 regularization. Our model is excessive in parameter, however, the performance does not change much when the weight decay penalty varies, as shown in Figure~\ref{figA_decay}.

\subsection{Ablation Studies on Positional Encoding}\label{secA7}

\begin{figure}[h]
\begin{center}
%\framebox[4.0in]{$\;$}
\includegraphics[height=6cm]{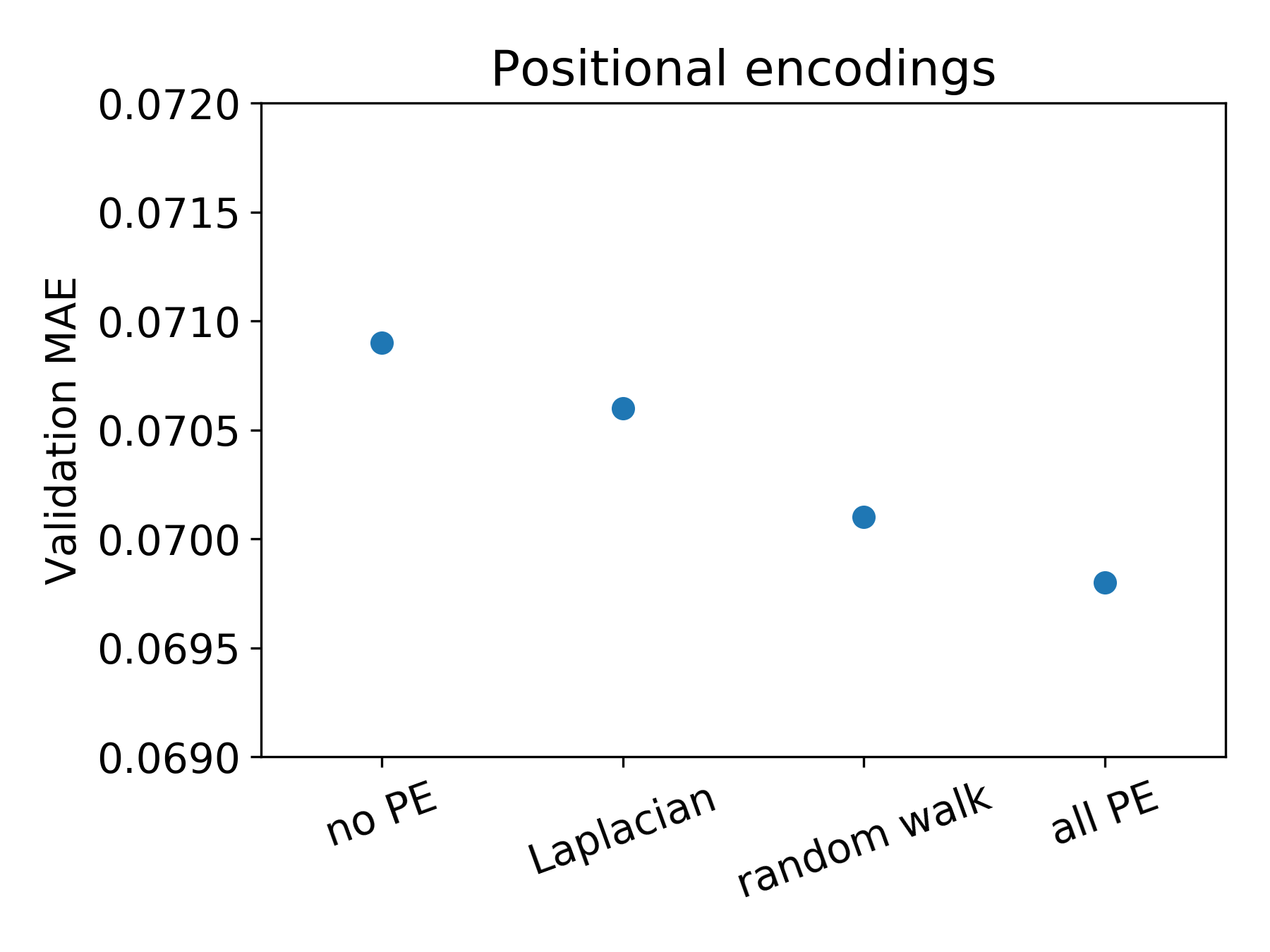}
\end{center}
\caption{Ablation studies on positional encoding before the GwT.} \label{figA_pe}
\end{figure}

The positional encoding for the graph is composed of two parts: Laplacian positional encoding and random walk positional encoding. As shown in Figure~\ref{figA_pe}, the Laplacian positional encoding and random walk positional encoding contribute in encoding the connectivity and structure of the graphs.

% \subsection{Correlation of Predictions and Ground Truth}\label{secA8}

% \begin{figure}[h]
% \begin{center}
% %\framebox[4.0in]{$\;$}
% \includegraphics[height=6cm]{./figures/corr_co2_s.png}
% \end{center}
% \caption{Scatter diagram of prediction and ground truth in task \texttt{co2\_adsorption}. x axis represents the label as the ground truth, y axis represents the predictions.} \label{figA_corr_co2}
% \end{figure}

% Figure~\ref{figA_corr_co2} illustrates the high accuracy of the CrysToGraph model's predictions, with all 5-fold test results plotted. The $r^2$ score for this task is calculated to be 0.8946. 

\subsection{Training Hyperparameters}\label{secA8}

The most ideal hyperparameter setting varies with different tasks. In Table~\ref{tabA_hyperparam}, we present a set of sample hyperparameters that was most widely applied on our training of models for benchmark tasks.

\begin{center}
\begin{table}[h]
\caption{Typical training hyperparameters of CrysToGraph models, some parameters may vary depends on 
specific task.}
\label{tabA_hyperparam}
\begin{tabular}{c|c}
\hline
\multicolumn{1}{c|}{\bf Names}  &\multicolumn{1}{c}{\bf Values}                     
\\ \hline 
eTGC blocks depth& 3\\
GwT layers depth& 1\\
FFNN layers depth& 1\\
projected feature dimension in each head& 32 or 24\\
number of head in attentions& 8\\
hidden feature dimension& 256\\
learning rate& 1e-4\\
optimizer& AdamW\\
weight decay& 1e-2 if overfitting else 0.0\\
batch size& 32\\
epochs& \makecell[c]{300, 600, 1000, 2000depends on dataset size}\\
ensemble size& 1\\
\hline

\end{tabular}
\end{table}
\end{center}

\section{Datasets and baseline models}

\subsection{Details of Traditional Crystal Datasets}\label{secA9}

We evaluated our model on 8 datasets of traditional crystals from the MatBench. The targets include a wide range from microscopic properties to macroscopic properties. The details of the datasets are shown in Table~\ref{tabA_matbench}. 
% The dummy predictions results of regression tasks are presented when all cross-validation outputs are assumed the mean of training test. For classification tasks, the dummy prediction is presented when cross-validation outputs is randomly sampled at the distribution of training set labels.
\begin{table}[h]
\caption{Details of the 8 datasets of MatBench, including target properties, number of samples and type of tasks.}\label{tabA_matbench}
% \begin{center}
% \resizebox{\textwidth}{!}{
\begin{tabular}{c|ccc}
\hline
\multicolumn{1}{c|}{\bf Datasets}  &\multicolumn{1}{c}{\bf Targets}  &\multicolumn{1}{c}{\bf Samples Size}        &\multicolumn{1}{c}{\bf Type of Task}           
\\ \hline 
\texttt{dielectric} & Refractive index (unitless) & 4,764& regression\\
\texttt{jdft2d} & Exfoliation energy ($\displaystyle meV/atom $) & 636& regression\\
\texttt{log\_gvrh} & \makecell[c]{Base 10 logarithm of the DFT Voigt-Reuss-Hill \\ average shear moduli in GPa} & 10,987& regression\\
\texttt{log\_kvrh} & \makecell[c]{Base 10 logarithm of the DFT Voigt-Reuss-Hill \\ average bulk moduli in GPa} & 10,987& regression\\
\texttt{mp\_e\_form} & Formation energy of bulk crystals ($\displaystyle eV/atom $)& 132,752& regression\\
\texttt{mp\_gap} & Band gap of bulk crystals ($\displaystyle eV $)& 106,113& regression\\
\texttt{\underline{mp\_is\_metal}}& Binary, 1 if structure is metal otherwise 0 & 106,113& classification\\
\texttt{phonons} & \makecell[c]{Frequency of the highest frequency \\ optical phonon mode peak ($\displaystyle cm^{-1} $)} & 1,265& regression\\
\hline

\end{tabular}
% }
% \end{center}
\end{table}

\subsection{Architecture of models on Benchmarks}\label{secA10}

The architecture of models on benchmark is compared in Table~\ref{tabA_models}. We listed some major components in the architecture in the table for a straightforward comparison. The components includes information of the general architecture, input data after pre-processing and the aggregation method for message passing in GNN.

\begin{center}
\begin{table}[h]
\caption{A comparison of architecture of models on benchmarks, presented with breaking them into major components.}
\label{tabA_models}
% \resizebox{\textwidth}{!}{
\begin{tabular}{c||cc|c|cc}
\hline
\multicolumn{1}{c||}{\bf Model}  &\multicolumn{1}{c}{\bf GNN}  &\multicolumn{1}{c|}{\bf Global Connection}  &\multicolumn{1}{c|}{\bf Line Graph} &\multicolumn{1}{c}{\bf Gated Aggregation} &\multicolumn{1}{c}{\bf Transformer Aggregation}\\ \hline 
MODNet& & \checkmark&  & &\\
CGCNN& \checkmark& &  & \checkmark&\\
ALIGNN& \checkmark& &  \checkmark& \checkmark&\\
coGN& \checkmark& &  & &\\
coNGN& \checkmark& &  \checkmark& &\\
Matformer& \checkmark& &  & &\checkmark\\
CrysToGraph& \checkmark& \checkmark&  \checkmark& &\checkmark\\
\hline

\end{tabular}
\end{table}
\end{center}

\end{appendices}